\documentclass[aps,prl,reprint]{revtex4-2}

\usepackage{graphicx}
\usepackage{textcomp}
\usepackage[version=3]{mhchem}
\usepackage{xcolor}

\usepackage[utf8]{inputenc}

\usepackage{lipsum}

\def\Ka{\ce{K\alpha}}
\def\Kaone{\ce{K\alpha_1}}
\def\NaMnO{\ce{NaMnO_4}}
\def\MnCl{\ce{MnCl_2}}

\begin{document}
\title{Observation of Multiplet Lines in Seeded Stimulated Mn K${\alpha}_1$ X-ray Emission}

\author{Thomas Kroll$^{1}$}
\email{tkroll@slac.stanford.edu}
\author{Margaret Doyle$^{2}$}
\author{Aliaksei Halavanau$^{3}$}
\author{Thomas M. Linker$^{4,5}$}
\author{Joshua Everts$^{1,6}$}
\author{Yurina Michine$^{7}$}
\author{Franklin D. Fuller$^{8}$}
\author{Clemens Weninger$^{9}$}
\author{Roberto Alonso-Mori$^{8}$}
\author{Claudio Pellegrini$^{3}$}
\author{Andrei Benediktovich$^{10}$}
\author{Makina Yabashi$^{11,12}$}
\author{Ichiro Inoue$^{11,14}$}
\author{Yuichi Inubushi$^{11,12}$}
\author{Taito Osaka$^{11}$}
\author{Toru Hara$^{11}$}
\author{Jumpei Yamada$^{11}$}
\author{Jan Kern$^{2}$}
\author{Junko Yano$^{2}$}
\author{Vittal K. Yachandra$^{2}$}
\author{Nina Rohringer$^{10,13}$}
\author{Hitoki Yoneda$^{7}$}
\author{Uwe Bergmann$^{4}$}
\email{ubergmann@wisc.edu}

\affiliation{$^{1}$ Stanford Synchrotron Radiation Lightsource, SLAC National Accelerator Laboratory, Menlo Park, CA 94025, USA}
\affiliation{$^{2}$ Molecular Biophysics and Integrated Bioimaging Division, Lawrence Berkeley National Laboratory, Berkeley, CA 94720, USA}
\affiliation{$^{3}$ Accelerator Research Division, SLAC National Accelerator Laboratory, Menlo Park, CA 94025, USA}
\affiliation{$^{4}$ Department of Physics, University of Wisconsin–Madison, Madison, WI 53706, USA}
\affiliation{$^{5}$ Stanford PULSE Institute, SLAC National Accelerator Laboratory, Menlo Park, CA 94025, USA}
\affiliation{$^{6}$ Department of Physics, University of Chicago, Chicago, IL 60637, USA}
\affiliation{$^{7}$ Institute for Laser Science, The University of Electro-Communications, Chofu,Tokyo 182-8585, Japan}
\affiliation{$^{8}$ Linac Coherent Light Source, SLAC National Accelerator Laboratory, Menlo Park, CA 94025, USA}
\affiliation{$^{9}$ MAX IV Laboratory, Lund University, Lund 224 84, Sweden}
\affiliation{$^{10}$ Center for Free-Electron Laser Science CFEL, Deutsches Elektronen-Synchrotron DESY, Hamburg 22603, Germany}
\affiliation{$^{11}$ RIKEN SPring-8 Center, Sayo-cho, Sayo-gun, Hyogo 679-5148, Japan}
\affiliation{$^{12}$ Japan Synchrotron Radiation Research Institute, Sayo-cho, Sayo-gun, Hyogo 679-5198, Japan}
\affiliation{$^{13}$ University of Hamburg, Institute for Experimental Physics/CFEL, 22761 Hamburg, Germany}
\affiliation{$^{14}$ Department of Physics, Universit{\"a}t Hamburg, Hamburg 20355, Germany}

\begin{abstract}
We report the successful resolution of the multiplet structure of the K${\alpha}_1$ x-ray emission in manganese (Mn) complexes through seeded stimulated X-ray emission spectroscopy (seeded S-XES). By employing a femtosecond pump pulse above the Mn K edge to generate simultaneous 1s core-holes, and a second-color tunable seed pulse to initiate the stimulated emission process, we were able to enhance individual lines within the K${\alpha}_1$ emission. This approach allows to resolve the fine multiplet features that are obscured by the life-time broadening in conventional Mn K${\alpha}$ XES. The work builds on our previous observation that S-XES from Mn(II) and Mn(VII) complexes pumped at high intensities can exhibit stimulated emission without sacrificing the chemical sensitivity to oxidation states. This technique opens the door to controlled high-resolution electronic structure spectroscopy in transition metal complexes beyond core hole life time broadening with potential applications in catalysis, inorganic chemistry, and materials science. 
\end{abstract}

\maketitle

X-ray emission spectroscopy (XES) is a powerful technique for determining the electronic structure of a large range of complexes, especially 3d transition metals \cite{Glatzel_CCR05, Vanko_PRB06, Zhang_Tracking_2014, Mara_Science17, Schuth_IC18}. The dependence of the K${\alpha}$ and K${\beta}$ XES spectral shape of 3d transition metal complexes is mainly given by the oxidation state, the spin state, and the covalent mixing strength\cite{Pollock_JACS14, Lafuerza_IC20, Guo_JPCA24} and other effects\cite{Lundberg_JACS13, Guo_PCCP16, Castillo_IC24}. Hence it is a powerful technique without the need for monochromatic or tunable incident x-rays. This advantage has made XES a popular technique at x-ray free-electron lasers (XFELs), where the spectrally broad self-amplified spontaneous emission (SASE) pulses can be used without monochromator. Furthermore, it can be simultaneously employed with x-ray scattering or diffraction \cite{Kern_Science13,Reinhard_NatureComm21}. An XES spectrum is comprised of several \ce{2p}$\rightarrow$\ce{1s} (K${\alpha}$) or \ce{3p}$\rightarrow$\ce{1s} (K${\beta}$) transitions from multiplet lines, and its overall spectral shape is determined by the individual strengths and energies of these lines. 
To quantitatively analyze XES spectra, different techniques including the integrated absolute difference (IAD) \cite{Vanko_PRB06}, the position of the first moment of the $\rm K\beta_{1,3}$ spectrum \cite{Messinger_JACS01, Lafuerza_IC20, Guo_JPCA24}, or the FWHM of the $\rm K\alpha_1$ spectrum \cite{Lafuerza_IC20} have been used. Simulating XES spectra with a theoretical code depends on the model used to describe the sample.
Generally, the main limitation of the sensitivity of K${\alpha}$ and K${\beta}$ XES spectrum to the electronic structure is the lack of being able to accurately measure the position or strengths of the multiplet lines due to the spectral broadening dominated by the short 1s core-hole life time. Hence, the fits often underestimate the number of transitions, especially those that are weak or close in energy.
In this study, we show how seeded stimulated X-ray emission spectroscopy (seeded S-XES)\cite{Yoneda_Atomic_2015, Kroll_PRL20, Doyle:23} can overcome this limitation and experimentally provide the energies of the Mn K${\alpha}_1$ multiplet transitions. 

\begin{figure}[htp]
 \centering
 \includegraphics[]{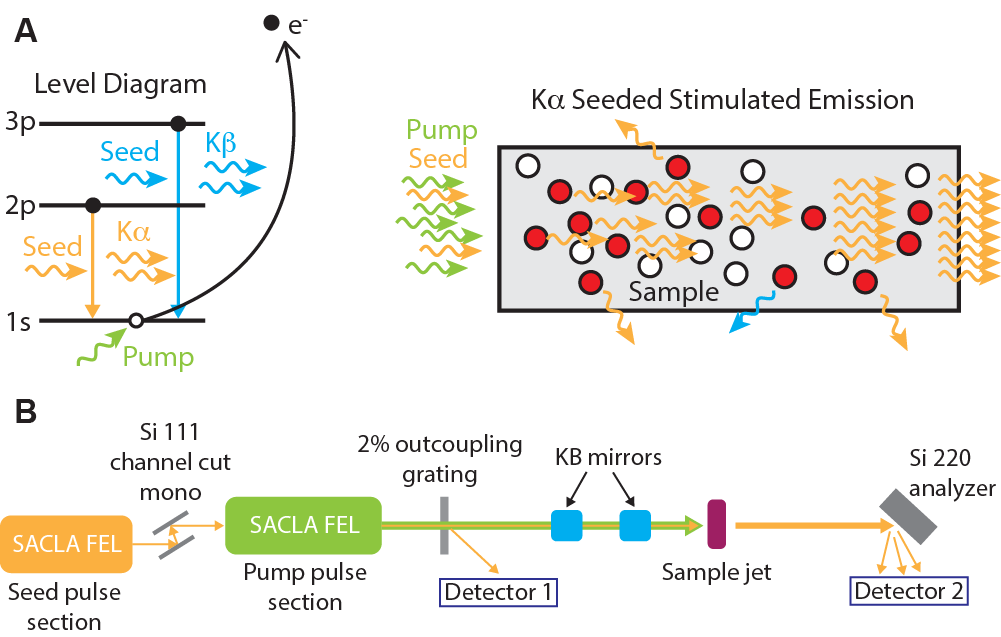}
 \caption{A: Sketch of the stimulated emission process including level diagram. B: Drawing of the experimental setup.}
 \label{fig:setup}
\end{figure}

S-XES is an inner-shell lasing process, where a short, highly focused, intense ($\sim10^{19}$ $\mathrm{W/cm^2}$) XFEL pulse \cite{Emma_First_2010, Ishikawa_compact_2012} tuned to a photon energy above the absorption edge creates a sufficiently large population of 1s core-hole excited states. Seeded by random photons out of the \ce{4\pi} solid angle spontaneous emission process that are along the direction of the excited state population, stimulated emission can emerge \cite{Rohringer_Atomic_2012, Beye_Stimulated_2013, Yoneda_Atomic_2015, Kroll_PRL18}. A sketch of this process is shown in Figure \ref{fig:setup}. This case is commonly known as amplified spontaneous emission (ASE) and when the collective emission time is short compared to the decoherence time it is also referred to as superfluorescence \cite{Benediktovitch_PRA19, Mercadier_PRL19}. We previously observed several effects when measuring the Mn \ce{K\alpha} ASE signal from two Mn solutions, \ce{NaMn(VII)O_4} and \ce{Mn(II)Cl_2}. The ASE spectra showed a) a chemical shift that is characteristic of the different Mn oxidation states, indicating a preservation of the chemical sensitivity; b) strong gain narrowing with spectral widths far below the 1s core-hole lifetime followed by broadening in the onset of the linear gain regime; and c) a broadening and a shift towards lower energies for the \ce{Mn(II)Cl_2} spectra in the saturation region \cite{Kroll_PRL18}. Because in ASE the strongest transition gets amplified before all others, the observed spectra showed mainly one strong peak. We speculated that the observed shift towards lower energies when reaching saturation in the \ce{Mn(II)Cl_2} ASE spectra indicates the onset of contributions from the weaker multiplet lines, known to be at lower photon energies. These ASE results showed the potential for this method for spectroscopy. However, without control over the seed in the stimulated emission processes, more detailed information about the electronic structure embedded in the K${\alpha}$ spectrum has not been accessible.

This limitation can be overcome by replacing the random spontaneous emission within the sample by an external tunable seed pulse that can be scanned across the photon energy of the multiplet lines that comprise the K${\alpha}$ spectrum. In this scheme, the pump pulse intensity is set below the threshold for ASE and stimulated emission only occurs when seeded by photons from the external pulse \cite{Yoneda_Atomic_2015, Doyle:23}. The schematics of seeded S-XES and the experimental setup is shown in Figure ~\ref{fig:setup} (See also Fig. 6 in ref \cite{Bergmann_PR24}.) We performed seeded S-XES experiments for two Mn complexes that have different electronic structures resulting in very different multiplet structures in their respective intermediate and final states: \ce{NaMn(VII)O_4}, which is formally a \ce{3d^0} complex exhibiting only one transition in its K${\alpha}_1$ spectrum, and \ce{MnCl(II)_2}, which is nominally a \ce{3d^5} high spin S=5/2 complex exhibiting five main transitions (and additional weaker ones) in its K${\alpha}_1$ spectrum as calculated by charge transfer multiplet simulations.\cite{Kroll_PRL18}

The experiments were performed at the SACLA XFEL facility in Japan in an experimental geometry described in Ref. \cite{Doyle:23}. We employed SACLA's two-color mode \cite{hara2013two,Inoue_JSR20}, which involves two sections of undulators, each set at different undulator strength parameter (K) values and separated by a magnetic chicane which detours the electron beam. One section generated the 6.6 keV pump pulse, and the other generated a tunable seed pulse tunable around the 5.9 keV Mn K${\alpha_1}$ photon energy.  A Si(111) channel-cut monochroator placed after the first set of undulators lead to a seed pulse bandwidth of ~0.7 eV FWHM with a Voigt line shape with ~20\% Lorentzian character as determined by a pseudo-Voigt fit. The tunable time delay between pump and seed pulse was set to 0 fs. Samples were delivered via a liquid jet with a diameter of 150 $\mu$m, and a flat Si(220) analyzer followed by a 2D detector (MPCCD) were used for obtaining information on the seeded S-XES signal. A second detector was placed upstream to analyze the seed pulse - thus providing information on the number of incoming seed photons, seed photon energy and seed pulse spectral shape. The schematics of this setup is shown in Figure \ref{fig:setup}B. 
To select the pump and seed pulse parameters for seeded S-XES, we first positioned the sample jet in focus at the maximum pump pulse intensity without employing a seed pulse. After observing ASE for \NaMnO, we moved the jet out of focus to reduce the pump pulse intensity until the ASE signal disappeared. This ensured that any S-XES signal would only arise in the presence of an external seed pulse. More details of the pump and seed pulse parameters for \NaMnO~ seeded S-XES at SACLA are described in Refs. \cite{Doyle:23} and \cite{Bergmann_PR24}. For \MnCl~ a slightly different protocol was used, as almost no ASE signal was observed at the pump pulse intensity employed. Here we set the seed pulse photon energy well below the Mn K${\alpha_1}$ line and optimized the focus until a seeded S-XES signal appeared. Thus, a small ASE contribution in the \MnCl~ seeded S-XES spectra cannot be excluded.


\begin{figure}[] 
 \centering
 \includegraphics[]{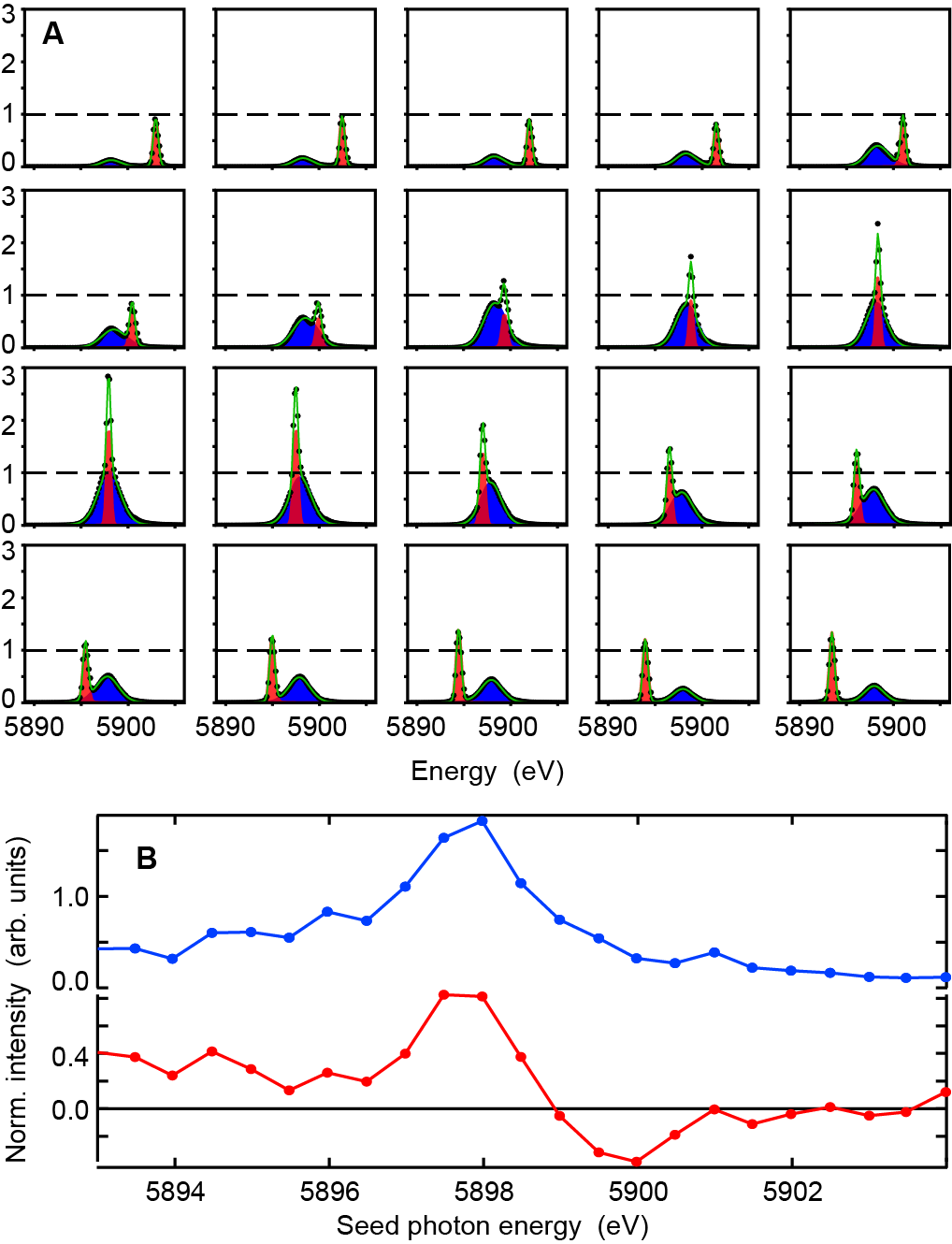}
 \caption{A: Collection of \NaMnO~ spectra for different seed pulses. The experimental data (filled circles) were fit with two pseudo-Voigt functions for the seed pulse (red) and the signal (blue). The vertical axes show the normalized intensity in arb. units.  B: Fit results versus the seed pulse photon energy. Blue: fit of signal, red: fit in seed pulse region minus the seed pulse signal as measured far away from the \Kaone~ line.}
 \label{fig:drive_through_NaMnO4}
\end{figure}

In Figure ~\ref{fig:drive_through_NaMnO4} A, a series of seeded S-XES spectra from \NaMnO~ is shown for a range of seed photon energies spanning $\sim$5 eV above and below the \Kaone~ energy region. Spectra are normalized by the seed pulse intensity measured in the upstream seed detector (Detector 1 in Figure \ref{fig:setup}B). The recorded spectra are shown as black circles fit by two pseudo-Voigt functions. The line shape of the seed photon spectrum (red) was fixed using the Voigt line shape with $\sim$20\% Lorentzian character in a constrained energy range given by the upstream detector. The signal intensity (blue) was fit freely.
An important observation is that already at a seed pulse photon energy $\sim$5 eV above the \Kaone~ peak position, a weak seeded S-XES signal at the \Kaone~ region appears (Figure ~\ref{fig:drive_through_NaMnO4} A top row). 

\par

Upon tuning the seed pulse photon energy closer to the \Kaone~ transition, the seeded S-XES signal slowly gains strength as more and more seed photons are present at the transition energy. When the seed pulse photon energy is at the \Kaone~ peak, the maximum seeded S-XES signal is reached. Detuning the seed pulse photon energy to lower energies results in a weaker the signal.
This is illustrated in Figure ~\ref{fig:drive_through_NaMnO4} B, where the fits of the signal are shown as a function of seed photon energy (blue curve). We note that the curve qualitatively follows the spontaneous XES spectrum with slightly higher intensity at lower energies. We assign this to stimulation of the very weak multiplet lines in this energy region.
Fits of the measured signal at the seed pulse spectral region minus the seed pulse signal (as measured far away from the \Kaone~ line) is shown in red. Two observations are made: (i) at higher energies close to \Kaone, the fitted narrow peak signal drops relative to the seed signal. This can be interpreted as the absorption of part of the seed pulse that does not temporally overlap with population inversion from states where the 2p electron has already refilled the 1s hole, leaving behind a 2p hole that lives for around 3 fs.\cite{Krause_1979} Instead of stimulating a \ce{2p}$\rightarrow$\ce{1s} decay, it triggers a \ce{1s}$\rightarrow$\ce{2p} excitation and gets absorbed more strongly.\cite{Weakly_NatureComm23} 
(ii) We observe gain narrowing as found by more photons in the spectral range of the seed pulse starting at energies close to the \Kaone~ transition.

\begin{figure}[bp]
 \centering
 \includegraphics[]{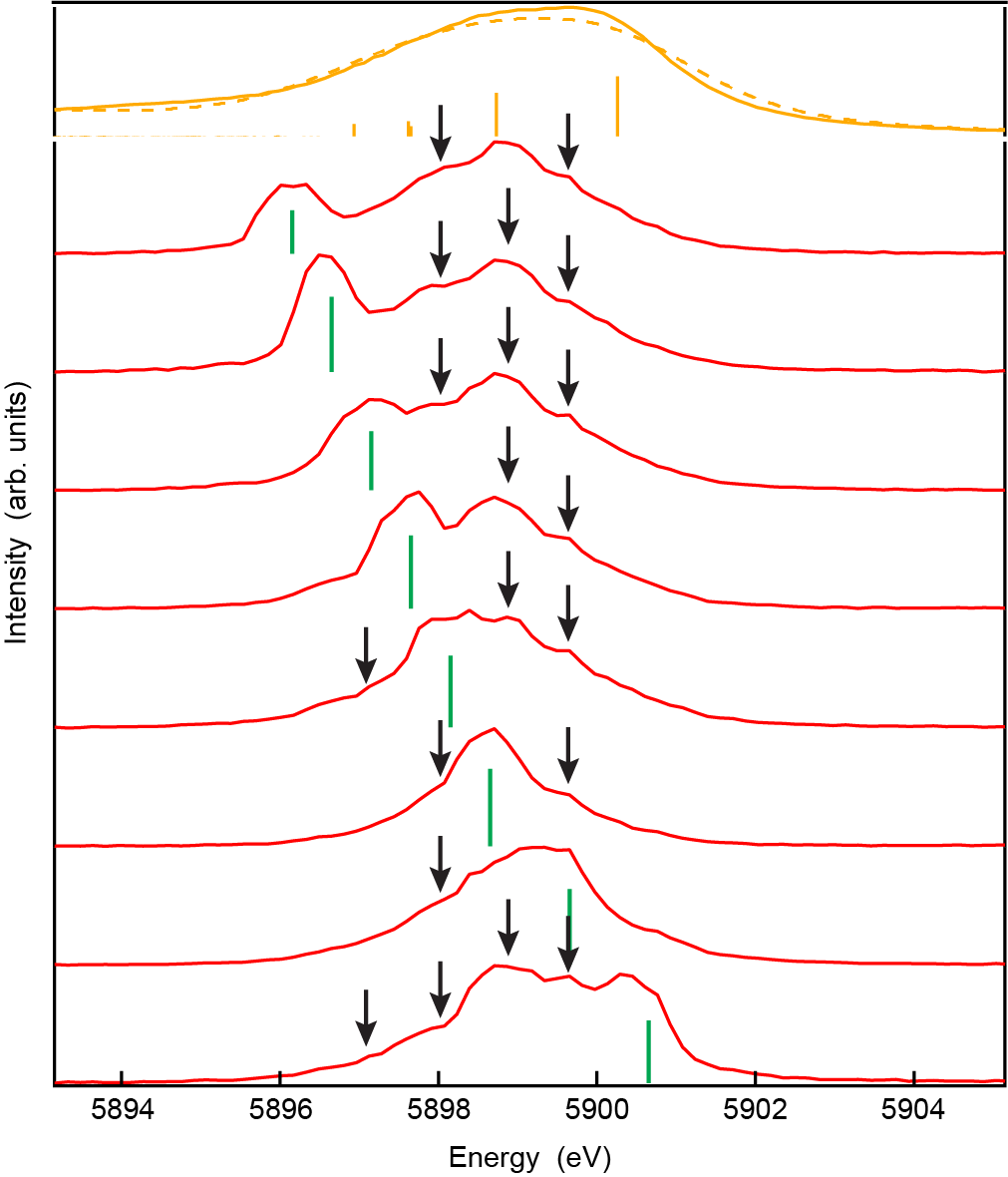}
 \caption{Experimental (solid orange) and theoretical (dashed orange) spontaneous XES spectra of \MnCl. The vertical lines represent the calculated individual transition intensities. Red curves are seeded stimulated XES spectra at a  given seed pulse photon energy (vertical green lines).}
 \label{fig:transitions_curve_MnCl2}
\end{figure}

Figure \ref{fig:transitions_curve_MnCl2} (top) shows the experimental \MnCl~ \Kaone~ spontaneous XES spectrum (dashed orange curve) together with the simulated spectrum (solid orange curve). The vertical orange lines represent the calculated multiplet transitions in  \MnCl.\cite{Kroll_PRL18} The red curves in Figure \ref{fig:transitions_curve_MnCl2} are seeded S-XES spectra at seed pulse photon energies represented by the vertical green lines. The spectra are obtained by averaging 4000 single-shot spectra at each seed pulse energy. Shown from top to bottom, the seed pulse is tuned to higher photon energies through the \Kaone~ region. 

Indicated by the black arrows are three spectral features at 5897.1, 5898.0, 5898.9, and 5899.7 eV. These features consistently appear in all cases, independent of the seed pulse energy. Their consistent appearance rules out that these features are the result of Poisson noise. With the following analysis we argue that these features represent the energies of the individual \Ka~ multiplet transitions. 

To show the existence and position of these additional features, we apply a peak finder algorithm that identifies peaks and their corresponding energies. It is based on a concept similar to that of a constant fraction discriminator (CFD). For this analysis, our raw signal was split into three components: One reference signal and two offset signals which were shifted in equal and opposite directions from the reference. A numpy.where() function (built-in feature of Python) was used to retrieve the indices corresponding to energies at which the reference signal was greater than either offset. Peaks were then assigned after selecting the local maximum from each identified cluster of points.  

\begin{figure}[t]
 \centering
 \includegraphics[]{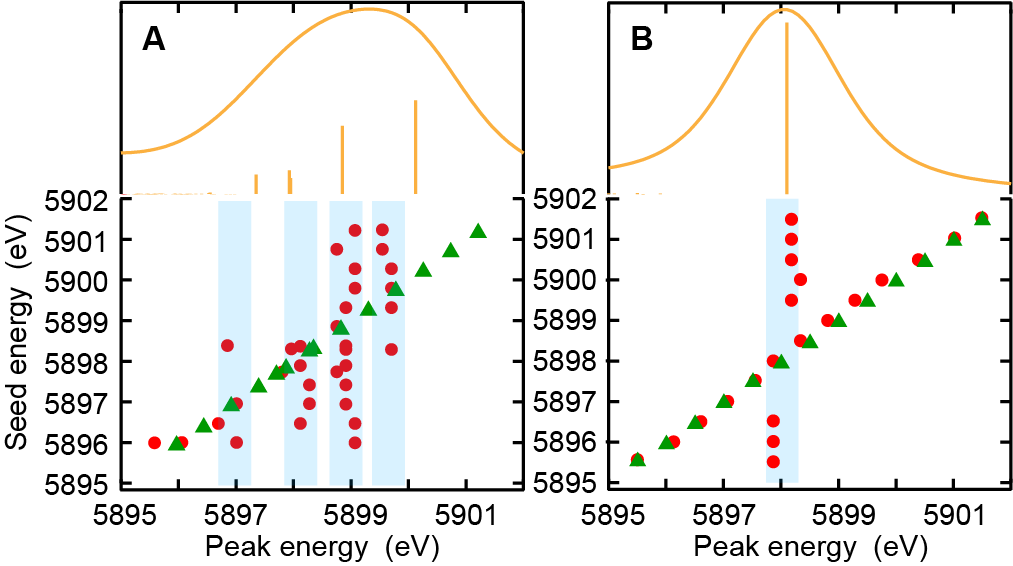}
 \caption{Results of the peak finder algorithm  together with the simulated spontaneous XES spectra for \MnCl~ (A) and \NaMnO~ (B).}
 \label{fig:peak_finder}
\end{figure}

The result for \MnCl~ is shown in Figure \ref{fig:peak_finder}A. For each seed energy (vertical axis) the results of the peak finder algorithm are plotted (horizontal axis). The green triangles show the results of the seed pulse photon energy as detected by the upstream detector and therefore appear on the diagonal. The red circles show the results of the peak finder algorithm. In addition to the peaks found on the diagonal (i.e. the seed pulse), additional peaks were found that appear at the same photon energy, independent of the seed pulse energy. From this analysis method, four energy regions are found that are highlighted in Figure \ref{fig:peak_finder} by light blue vertical bars as guide to the eye. The four distinguishable transitions agree with the theoretical result. Note that at 5898 eV two transitions are close to each other that cannot be distinguished experimentally. We also find a considerable difference between the energy positions of the calculated (yellow) and measured (red) transitions. Another striking observation is the fact that the first transition, calculated as the strongest transition, does not lead to the strongest intensity (see spectra in Figure \ref{fig:transitions_curve_MnCl2}). This suggests that either the multiplet calculation result does not find the correct intensities/energies, or that in seeded S-XES the strongest peak does not arise from the strongest individual transition, but (as in spontaneous emission) from the overlap of two or more individual transitions that add up to the strongest spectral feature. As this depends on the spectral width and intensity of the seed pulse, this can in principle be controlled with better seed pulse optimization.   

\begin{figure}[t]
 \centering
 \includegraphics[width=0.5\textwidth]{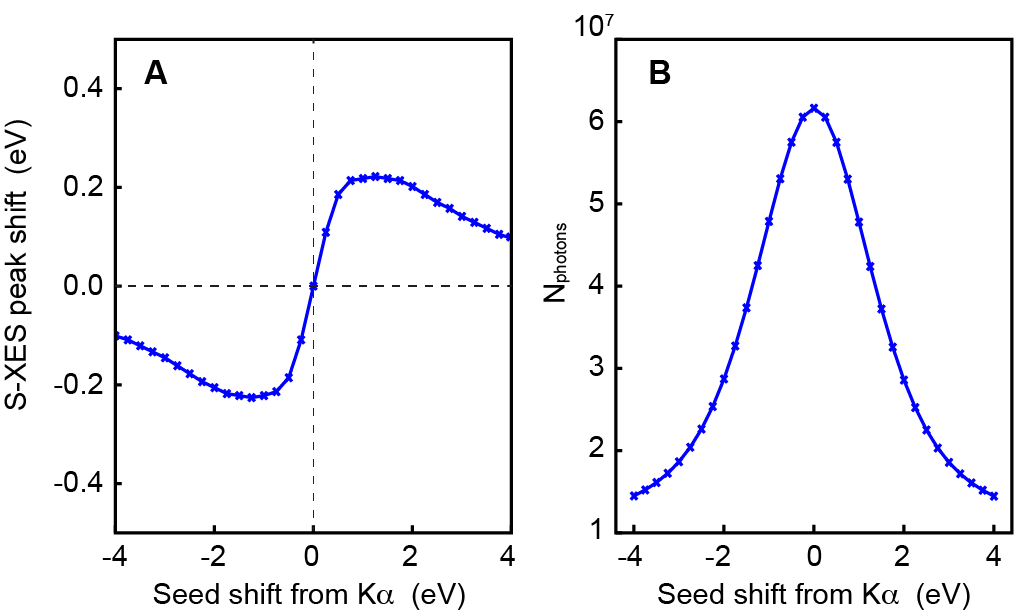}
 \caption{(A) Calculated spectral shift in SS-XES for different seed photon energies for a single transition line. (B) Calculated number of stimulated photons as function of seed pulse energy. }
 \label{fig:coupled_oscillator}
\end{figure}

To further check our finding of distinct transition energies in seeded S-XES from \MnCl~ and our claim that these are related to the multiplet lines, we also measured \NaMnO~ measured as reference. \NaMnO~ has a nominal oxidation state of +7 and hence a formal \ce{3d^0} configuration. 
The results of the peak finder algorithm for the seeded stimulated XES experiment on \NaMnO~ is shown in Figure \ref{fig:peak_finder}B. Only one peak (besides that of the seed pulse) is found at $\sim$5898 eV as highlighted by the vertical light blue bar. This result agrees with the theory, which shows only one transition in the Mn \Kaone~ spectrum in a \ce{3d^0} configuration. \cite{Kroll_PRL18} This result corroborates that the multiple peaks found for \MnCl seeded S-XES are the result of the more much complex \Kaone~ mutiplet structure in \MnCl~ as compared to \NaMnO. 

An interesting phenomenon in the \NaMnO~ seeded S-XES peaks shown in Figure \ref{fig:peak_finder}B is that the energy position of the transitions shifts to slightly lower energies when the seed pulse crosses to energies below the \Ka\ce{_1} transition. We attribute this behavior to a coupled oscillator effect that pulls the stimulated emission signal towards the seed pulse energy, which is a well know phenomena in both the laser and FEL communities. \cite{Fader_IEEE85, Yaryv_book89, Allaria_EPL10}
To quantify these effects, we show the results from 3D Maxwell Bloch simulations \cite{Chuchurka_PRA24, Linker_arXiv24} in  Figure \ref{fig:coupled_oscillator}, where we simulate the seeded S-XES process initiated by a weak SASE spike in the pump pulse and a seed pulse with a Lorentzian spectral profile. 
In line to the experiment, Figure \ref{fig:coupled_oscillator}A illustrates that the position of the \Kaone~ signal shifts when the seed pulse energy crosses the transition energy with a maximum pulling of ${\approx}$0.2 eV. While the spectra shift with the seed pulse, the number of stimulated photons is always highest when the seed is resonant with transition, which is illustrated in Figure \ref{fig:coupled_oscillator}B. In contrast, we did not observe frequency pulling for \MnCl~ \Kaone~ seeded S-XES. We note that the observed and simulated shift due to frequency pulling for a single transition in \NaMnO~ is only ${\approx}$0.2 eV and might not be discernible in our date. The simulations also show that the coupling is strongest when the seed is within ${\approx}$1 eV from the transition. Thus, the lack of observed frequency pulling in the \MnCl~ \Kaone~ signal is most likely due to the fact that the internal coupling between the multiplet resonances out competes any frequency pulling by the seed pulse.

In conclusion, we have shown that seeded S-XES can provide direct spectral information of the multiplet structure underlying the Mn \Kaone~ X-ray emission in two Mn compounds with different electronic structure. As shown in our recent work, this approach is feasible even for systems with low metal concentrations, possibly metalloproteins. \cite{Doyle:23, Bergmann_PR24}. The main challenge of this technique with our current experimental setup is the limit in seeding control. This is evidenced by the observed \Kaone~ seeded S-XES signal, even when the seed pulse is far away from the resonance. Control of the seed pulse intensity independent from that of the pump pulse can overcome this limitation. This allows to optimize the parameters such, that seeded S-XES only occurs when the seed is very close to a resonance. 

Acknowledgment: We thank Kirk Larsen for providing us with the peak finder algorithm. This work is supported by the U.S. Department of Energy, Office of Science, Basic Energy Sciences, under Awards DE-SC0023270 to U.B. and DE-AC02-76SF00515 to A.H., C.P., U.B.; by the Director, Office of Science, Office of Basic Energy Sciences (OBES), Division of Chemical Sciences, Geosciences, and Biosciences of the Department of Energy (DOE) (J.Y., V.K.Y. and J.K.); Laboratory Directed Research and Development Program at SLAC National Accelerator Laboratory (DE-AC02-76SF00515 to U.B.); the Ruth L. Kirschstein National Research Service Award (F32GM116423 to F.D.F.); NIH (GM126289 to J.K., GM110501 to J.Y., GM149528 to V.K.Y.); and JPSJ KAKENHI and JST PRESTO. The Stanford Synchrotron Radiation Lightsource Structural Molecular Biology Program is supported by the DOE Office of Biological and
Environmental Research and NIH National Institute of General Medical Sciences (NIGMS). The contents of this publication are solely the responsibility of the
authors and do not necessarily represent the official views of the NIGMS or the NIH. The experiment at SACLA was performed with the approval of the Japan Synchrotron Radiation Research Institute (proposal no. 2021B8075).

\bibliographystyle{apsrev4-2}
\bibliography{references} 

\begin{thebibliography}{34}%
\makeatletter
\providecommand \@ifxundefined [1]{%
 \@ifx{#1\undefined}
}%
\providecommand \@ifnum [1]{%
 \ifnum #1\expandafter \@firstoftwo
 \else \expandafter \@secondoftwo
 \fi
}%
\providecommand \@ifx [1]{%
 \ifx #1\expandafter \@firstoftwo
 \else \expandafter \@secondoftwo
 \fi
}%
\providecommand \natexlab [1]{#1}%
\providecommand \enquote  [1]{``#1''}%
\providecommand \bibnamefont  [1]{#1}%
\providecommand \bibfnamefont [1]{#1}%
\providecommand \citenamefont [1]{#1}%
\providecommand \href@noop [0]{\@secondoftwo}%
\providecommand \href [0]{\begingroup \@sanitize@url \@href}%
\providecommand \@href[1]{\@@startlink{#1}\@@href}%
\providecommand \@@href[1]{\endgroup#1\@@endlink}%
\providecommand \@sanitize@url [0]{\catcode `\\12\catcode `\$12\catcode `\&12\catcode `\#12\catcode `\^12\catcode `\_12\catcode `\%12\relax}%
\providecommand \@@startlink[1]{}%
\providecommand \@@endlink[0]{}%
\providecommand \url  [0]{\begingroup\@sanitize@url \@url }%
\providecommand \@url [1]{\endgroup\@href {#1}{\urlprefix }}%
\providecommand \urlprefix  [0]{URL }%
\providecommand \Eprint [0]{\href }%
\providecommand \doibase [0]{https://doi.org/}%
\providecommand \selectlanguage [0]{\@gobble}%
\providecommand \bibinfo  [0]{\@secondoftwo}%
\providecommand \bibfield  [0]{\@secondoftwo}%
\providecommand \translation [1]{[#1]}%
\providecommand \BibitemOpen [0]{}%
\providecommand \bibitemStop [0]{}%
\providecommand \bibitemNoStop [0]{.\EOS\space}%
\providecommand \EOS [0]{\spacefactor3000\relax}%
\providecommand \BibitemShut  [1]{\csname bibitem#1\endcsname}%
\let\auto@bib@innerbib\@empty
\bibitem [{\citenamefont {Glatzel}\ and\ \citenamefont {Bergmann}(2005)}]{Glatzel_CCR05}%
  \BibitemOpen
  \bibfield  {author} {\bibinfo {author} {\bibfnamefont {P.}~\bibnamefont {Glatzel}}\ and\ \bibinfo {author} {\bibfnamefont {U.}~\bibnamefont {Bergmann}},\ }\href@noop {} {\bibfield  {journal} {\bibinfo  {journal} {Coord. Chem. Rev.}\ }\textbf {\bibinfo {volume} {249}},\ \bibinfo {pages} {65} (\bibinfo {year} {2005})}\BibitemShut {NoStop}%
\bibitem [{\citenamefont {Vanko}\ \emph {et~al.}(2006)\citenamefont {Vanko}, \citenamefont {Rueff}, \citenamefont {Mattila}, \citenamefont {Nemeth},\ and\ \citenamefont {Shukla}}]{Vanko_PRB06}%
  \BibitemOpen
  \bibfield  {author} {\bibinfo {author} {\bibfnamefont {G.}~\bibnamefont {Vanko}}, \bibinfo {author} {\bibfnamefont {J.-P.}\ \bibnamefont {Rueff}}, \bibinfo {author} {\bibfnamefont {A.}~\bibnamefont {Mattila}}, \bibinfo {author} {\bibfnamefont {Z.}~\bibnamefont {Nemeth}},\ and\ \bibinfo {author} {\bibfnamefont {A.}~\bibnamefont {Shukla}},\ }\href@noop {} {\bibfield  {journal} {\bibinfo  {journal} {Phys. Rev. B}\ }\textbf {\bibinfo {volume} {73}},\ \bibinfo {pages} {024424} (\bibinfo {year} {2006})}\BibitemShut {NoStop}%
\bibitem [{\citenamefont {Zhang}\ \emph {et~al.}(2014)\citenamefont {Zhang}, \citenamefont {Alonso-Mori}, \citenamefont {Bergmann}, \citenamefont {Bressler}, \citenamefont {Chollet}, \citenamefont {Galler}, \citenamefont {Gawelda}, \citenamefont {Hadt}, \citenamefont {Hartsock}, \citenamefont {Kroll}, \citenamefont {Kj\ae{}r}, \citenamefont {Kubi{\v c}ek}, \citenamefont {Lemke}, \citenamefont {Liang}, \citenamefont {Meyer}, \citenamefont {Nielsen}, \citenamefont {Purser}, \citenamefont {Robinson}, \citenamefont {Solomon}, \citenamefont {Sun}, \citenamefont {Sokaras}, \citenamefont {{van Driel}}, \citenamefont {Vank{\'o}}, \citenamefont {Weng}, \citenamefont {Zhu},\ and\ \citenamefont {Gaffney}}]{Zhang_Tracking_2014}%
  \BibitemOpen
  \bibfield  {author} {\bibinfo {author} {\bibfnamefont {W.}~\bibnamefont {Zhang}}, \bibinfo {author} {\bibfnamefont {R.}~\bibnamefont {Alonso-Mori}}, \bibinfo {author} {\bibfnamefont {U.}~\bibnamefont {Bergmann}}, \bibinfo {author} {\bibfnamefont {C.}~\bibnamefont {Bressler}}, \bibinfo {author} {\bibfnamefont {M.}~\bibnamefont {Chollet}}, \bibinfo {author} {\bibfnamefont {A.}~\bibnamefont {Galler}}, \bibinfo {author} {\bibfnamefont {W.}~\bibnamefont {Gawelda}}, \bibinfo {author} {\bibfnamefont {R.~G.}\ \bibnamefont {Hadt}}, \bibinfo {author} {\bibfnamefont {R.~W.}\ \bibnamefont {Hartsock}}, \bibinfo {author} {\bibfnamefont {T.}~\bibnamefont {Kroll}}, \bibinfo {author} {\bibfnamefont {K.~S.}\ \bibnamefont {Kj\ae{}r}}, \bibinfo {author} {\bibfnamefont {K.}~\bibnamefont {Kubi{\v c}ek}}, \bibinfo {author} {\bibfnamefont {H.~T.}\ \bibnamefont {Lemke}}, \bibinfo {author} {\bibfnamefont {H.~W.}\ \bibnamefont {Liang}}, \bibinfo {author} {\bibfnamefont {D.~A.}\ \bibnamefont {Meyer}}, \bibinfo {author} {\bibfnamefont
  {M.~M.}\ \bibnamefont {Nielsen}}, \bibinfo {author} {\bibfnamefont {C.}~\bibnamefont {Purser}}, \bibinfo {author} {\bibfnamefont {J.~S.}\ \bibnamefont {Robinson}}, \bibinfo {author} {\bibfnamefont {E.~I.}\ \bibnamefont {Solomon}}, \bibinfo {author} {\bibfnamefont {Z.}~\bibnamefont {Sun}}, \bibinfo {author} {\bibfnamefont {D.}~\bibnamefont {Sokaras}}, \bibinfo {author} {\bibfnamefont {T.~B.}\ \bibnamefont {{van Driel}}}, \bibinfo {author} {\bibfnamefont {G.}~\bibnamefont {Vank{\'o}}}, \bibinfo {author} {\bibfnamefont {T.-C.}\ \bibnamefont {Weng}}, \bibinfo {author} {\bibfnamefont {D.}~\bibnamefont {Zhu}},\ and\ \bibinfo {author} {\bibfnamefont {K.~J.}\ \bibnamefont {Gaffney}},\ }\href {https://doi.org/10.1038/nature13252} {\bibfield  {journal} {\bibinfo  {journal} {Nature}\ }\textbf {\bibinfo {volume} {509}},\ \bibinfo {pages} {345} (\bibinfo {year} {2014})}\BibitemShut {NoStop}%
\bibitem [{\citenamefont {Mara}\ \emph {et~al.}(2017)\citenamefont {Mara}, \citenamefont {Hadt}, \citenamefont {Reinhard}, \citenamefont {Kroll}, \citenamefont {Lim}, \citenamefont {Hartsock}, \citenamefont {Alonso-Mori}, \citenamefont {Chollet}, \citenamefont {Glownia}, \citenamefont {Nelson}, \citenamefont {Sokaras}, \citenamefont {Kunnus}, \citenamefont {Hodgson}, \citenamefont {Hedman}, \citenamefont {Bergmann}, \citenamefont {Gaffney},\ and\ \citenamefont {Solomon}}]{Mara_Science17}%
  \BibitemOpen
  \bibfield  {author} {\bibinfo {author} {\bibfnamefont {M.~W.}\ \bibnamefont {Mara}}, \bibinfo {author} {\bibfnamefont {R.~G.}\ \bibnamefont {Hadt}}, \bibinfo {author} {\bibfnamefont {M.~E.}\ \bibnamefont {Reinhard}}, \bibinfo {author} {\bibfnamefont {T.}~\bibnamefont {Kroll}}, \bibinfo {author} {\bibfnamefont {H.}~\bibnamefont {Lim}}, \bibinfo {author} {\bibfnamefont {R.~W.}\ \bibnamefont {Hartsock}}, \bibinfo {author} {\bibfnamefont {R.}~\bibnamefont {Alonso-Mori}}, \bibinfo {author} {\bibfnamefont {M.}~\bibnamefont {Chollet}}, \bibinfo {author} {\bibfnamefont {J.~M.}\ \bibnamefont {Glownia}}, \bibinfo {author} {\bibfnamefont {S.}~\bibnamefont {Nelson}}, \bibinfo {author} {\bibfnamefont {D.}~\bibnamefont {Sokaras}}, \bibinfo {author} {\bibfnamefont {K.}~\bibnamefont {Kunnus}}, \bibinfo {author} {\bibfnamefont {K.~O.}\ \bibnamefont {Hodgson}}, \bibinfo {author} {\bibfnamefont {B.}~\bibnamefont {Hedman}}, \bibinfo {author} {\bibfnamefont {U.}~\bibnamefont {Bergmann}}, \bibinfo {author} {\bibfnamefont
  {K.~J.}\ \bibnamefont {Gaffney}},\ and\ \bibinfo {author} {\bibfnamefont {E.~I.}\ \bibnamefont {Solomon}},\ }\href@noop {} {\bibfield  {journal} {\bibinfo  {journal} {Science}\ }\textbf {\bibinfo {volume} {356}},\ \bibinfo {pages} {1276} (\bibinfo {year} {2017})}\BibitemShut {NoStop}%
\bibitem [{\citenamefont {Schuth}\ \emph {et~al.}(2018)\citenamefont {Schuth}, \citenamefont {Zaharieva}, \citenamefont {Chernev}, \citenamefont {Berggren}, \citenamefont {Anderlund}, \citenamefont {Styring}, \citenamefont {Dau},\ and\ \citenamefont {Haumann}}]{Schuth_IC18}%
  \BibitemOpen
  \bibfield  {author} {\bibinfo {author} {\bibfnamefont {N.}~\bibnamefont {Schuth}}, \bibinfo {author} {\bibfnamefont {I.}~\bibnamefont {Zaharieva}}, \bibinfo {author} {\bibfnamefont {P.}~\bibnamefont {Chernev}}, \bibinfo {author} {\bibfnamefont {G.}~\bibnamefont {Berggren}}, \bibinfo {author} {\bibfnamefont {M.}~\bibnamefont {Anderlund}}, \bibinfo {author} {\bibfnamefont {S.}~\bibnamefont {Styring}}, \bibinfo {author} {\bibfnamefont {H.}~\bibnamefont {Dau}},\ and\ \bibinfo {author} {\bibfnamefont {M.}~\bibnamefont {Haumann}},\ }\href@noop {} {\bibfield  {journal} {\bibinfo  {journal} {Inorganic Chemistry}\ }\textbf {\bibinfo {volume} {57}},\ \bibinfo {pages} {10424} (\bibinfo {year} {2018})}\BibitemShut {NoStop}%
\bibitem [{\citenamefont {Pollock}\ \emph {et~al.}(2014)\citenamefont {Pollock}, \citenamefont {Delgado-Jaime}, \citenamefont {Atanasov}, \citenamefont {Neese},\ and\ \citenamefont {DeBeer}}]{Pollock_JACS14}%
  \BibitemOpen
  \bibfield  {author} {\bibinfo {author} {\bibfnamefont {C.~J.}\ \bibnamefont {Pollock}}, \bibinfo {author} {\bibfnamefont {M.~U.}\ \bibnamefont {Delgado-Jaime}}, \bibinfo {author} {\bibfnamefont {M.}~\bibnamefont {Atanasov}}, \bibinfo {author} {\bibfnamefont {F.}~\bibnamefont {Neese}},\ and\ \bibinfo {author} {\bibfnamefont {S.}~\bibnamefont {DeBeer}},\ }\href@noop {} {\bibfield  {journal} {\bibinfo  {journal} {J. Am. Chem. Soc.}\ }\textbf {\bibinfo {volume} {136}},\ \bibinfo {pages} {9453} (\bibinfo {year} {2014})}\BibitemShut {NoStop}%
\bibitem [{\citenamefont {Lafuerza}\ \emph {et~al.}(2020)\citenamefont {Lafuerza}, \citenamefont {Carlantuono}, \citenamefont {Retegan},\ and\ \citenamefont {Glatzel}}]{Lafuerza_IC20}%
  \BibitemOpen
  \bibfield  {author} {\bibinfo {author} {\bibfnamefont {S.}~\bibnamefont {Lafuerza}}, \bibinfo {author} {\bibfnamefont {A.}~\bibnamefont {Carlantuono}}, \bibinfo {author} {\bibfnamefont {M.}~\bibnamefont {Retegan}},\ and\ \bibinfo {author} {\bibfnamefont {P.}~\bibnamefont {Glatzel}},\ }\href@noop {} {\bibfield  {journal} {\bibinfo  {journal} {Inorg. Chem.}\ }\textbf {\bibinfo {volume} {59}},\ \bibinfo {pages} {12518} (\bibinfo {year} {2020})}\BibitemShut {NoStop}%
\bibitem [{\citenamefont {Lafuerza}\ \emph {et~al.}(2024)\citenamefont {Lafuerza}, \citenamefont {Carlantuono}, \citenamefont {Retegan},\ and\ \citenamefont {Glatzel}}]{Guo_JPCA24}%
  \BibitemOpen
  \bibfield  {author} {\bibinfo {author} {\bibfnamefont {S.}~\bibnamefont {Lafuerza}}, \bibinfo {author} {\bibfnamefont {A.}~\bibnamefont {Carlantuono}}, \bibinfo {author} {\bibfnamefont {M.}~\bibnamefont {Retegan}},\ and\ \bibinfo {author} {\bibfnamefont {P.}~\bibnamefont {Glatzel}},\ }\href@noop {} {\bibfield  {journal} {\bibinfo  {journal} {Jour. Phys. Chem. A}\ }\textbf {\bibinfo {volume} {128}},\ \bibinfo {pages} {1260} (\bibinfo {year} {2024})}\BibitemShut {NoStop}%
\bibitem [{\citenamefont {Lundberg}\ \emph {et~al.}(2013)\citenamefont {Lundberg}, \citenamefont {Kroll}, \citenamefont {DeBeer}, \citenamefont {Bergmann}, \citenamefont {Wilson}, \citenamefont {Glatzel}, \citenamefont {Nordlund}, \citenamefont {Hedman}, \citenamefont {Hodgson},\ and\ \citenamefont {Solomon}}]{Lundberg_JACS13}%
  \BibitemOpen
  \bibfield  {author} {\bibinfo {author} {\bibfnamefont {M.}~\bibnamefont {Lundberg}}, \bibinfo {author} {\bibfnamefont {T.}~\bibnamefont {Kroll}}, \bibinfo {author} {\bibfnamefont {S.}~\bibnamefont {DeBeer}}, \bibinfo {author} {\bibfnamefont {U.}~\bibnamefont {Bergmann}}, \bibinfo {author} {\bibfnamefont {S.~A.}\ \bibnamefont {Wilson}}, \bibinfo {author} {\bibfnamefont {P.}~\bibnamefont {Glatzel}}, \bibinfo {author} {\bibfnamefont {D.}~\bibnamefont {Nordlund}}, \bibinfo {author} {\bibfnamefont {B.}~\bibnamefont {Hedman}}, \bibinfo {author} {\bibfnamefont {K.~O.}\ \bibnamefont {Hodgson}},\ and\ \bibinfo {author} {\bibfnamefont {E.~I.}\ \bibnamefont {Solomon}},\ }\href@noop {} {\bibfield  {journal} {\bibinfo  {journal} {J. Am. Chem. Soc.}\ }\textbf {\bibinfo {volume} {135}},\ \bibinfo {pages} {17121} (\bibinfo {year} {2013})}\BibitemShut {NoStop}%
\bibitem [{\citenamefont {Guo}\ \emph {et~al.}(2016)\citenamefont {Guo}, \citenamefont {S{\o}rensen}, \citenamefont {Delcey}, \citenamefont {Pinjari},\ and\ \citenamefont {Lundberg}}]{Guo_PCCP16}%
  \BibitemOpen
  \bibfield  {author} {\bibinfo {author} {\bibfnamefont {M.}~\bibnamefont {Guo}}, \bibinfo {author} {\bibfnamefont {L.~K.}\ \bibnamefont {S{\o}rensen}}, \bibinfo {author} {\bibfnamefont {M.~G.}\ \bibnamefont {Delcey}}, \bibinfo {author} {\bibfnamefont {R.~V.}\ \bibnamefont {Pinjari}},\ and\ \bibinfo {author} {\bibfnamefont {M.}~\bibnamefont {Lundberg}},\ }\href@noop {} {\bibfield  {journal} {\bibinfo  {journal} {Phys. Chem. Chem. Phys.}\ }\textbf {\bibinfo {volume} {18}},\ \bibinfo {pages} {3250} (\bibinfo {year} {2016})}\BibitemShut {NoStop}%
\bibitem [{\citenamefont {Castillo}\ \emph {et~al.}(2024)\citenamefont {Castillo}, \citenamefont {Kuiken}, \citenamefont {Weyherm{\"u}ller},\ and\ \citenamefont {DeBeer}}]{Castillo_IC24}%
  \BibitemOpen
  \bibfield  {author} {\bibinfo {author} {\bibfnamefont {R.~G.}\ \bibnamefont {Castillo}}, \bibinfo {author} {\bibfnamefont {B.~E.~V.}\ \bibnamefont {Kuiken}}, \bibinfo {author} {\bibfnamefont {T.}~\bibnamefont {Weyherm{\"u}ller}},\ and\ \bibinfo {author} {\bibfnamefont {S.}~\bibnamefont {DeBeer}},\ }\href@noop {} {\bibfield  {journal} {\bibinfo  {journal} {Inorganic Chemistry}\ }\textbf {\bibinfo {volume} {63}},\ \bibinfo {pages} {18468} (\bibinfo {year} {2024})}\BibitemShut {NoStop}%
\bibitem [{\citenamefont {Kern}\ \emph {et~al.}(2013)\citenamefont {Kern}, \citenamefont {Alonso-Mori}, \citenamefont {Tran}, \citenamefont {Hattne}, \citenamefont {Gildea}, \citenamefont {Echols}, \citenamefont {Gl{\"o}ckner}, \citenamefont {Hellmich}, \citenamefont {Laksmono}, \citenamefont {Sierra}, \citenamefont {Lassalle-Kaiser}, \citenamefont {Koroidov}, \citenamefont {Lampe}, \citenamefont {Han}, \citenamefont {Gul}, \citenamefont {DiFiore}, \citenamefont {Milathianaki}, \citenamefont {Fry}, \citenamefont {Miahnahri}, \citenamefont {Schafer}, \citenamefont {Messerschmidt}, \citenamefont {Seibert}, \citenamefont {Koglin}, \citenamefont {Sokaras}, \citenamefont {Weng}, \citenamefont {Sellberg}, \citenamefont {Latimer}, \citenamefont {Grosse-Kunstleve}, \citenamefont {Zwart}, \citenamefont {White}, \citenamefont {Glatzel}, \citenamefont {Adams}, \citenamefont {Bogan}, \citenamefont {Williams}, \citenamefont {Boutet}, \citenamefont {Messinger}, \citenamefont {Zouni}, \citenamefont {Sauter}, \citenamefont
  {Yachandra}, \citenamefont {Bergmann},\ and\ \citenamefont {Yano}}]{Kern_Science13}%
  \BibitemOpen
  \bibfield  {author} {\bibinfo {author} {\bibfnamefont {J.}~\bibnamefont {Kern}}, \bibinfo {author} {\bibfnamefont {R.}~\bibnamefont {Alonso-Mori}}, \bibinfo {author} {\bibfnamefont {R.}~\bibnamefont {Tran}}, \bibinfo {author} {\bibfnamefont {J.}~\bibnamefont {Hattne}}, \bibinfo {author} {\bibfnamefont {R.~J.}\ \bibnamefont {Gildea}}, \bibinfo {author} {\bibfnamefont {N.}~\bibnamefont {Echols}}, \bibinfo {author} {\bibfnamefont {C.}~\bibnamefont {Gl{\"o}ckner}}, \bibinfo {author} {\bibfnamefont {J.}~\bibnamefont {Hellmich}}, \bibinfo {author} {\bibfnamefont {H.}~\bibnamefont {Laksmono}}, \bibinfo {author} {\bibfnamefont {R.~G.}\ \bibnamefont {Sierra}}, \bibinfo {author} {\bibfnamefont {B.}~\bibnamefont {Lassalle-Kaiser}}, \bibinfo {author} {\bibfnamefont {S.}~\bibnamefont {Koroidov}}, \bibinfo {author} {\bibfnamefont {A.}~\bibnamefont {Lampe}}, \bibinfo {author} {\bibfnamefont {G.}~\bibnamefont {Han}}, \bibinfo {author} {\bibfnamefont {S.}~\bibnamefont {Gul}}, \bibinfo {author} {\bibfnamefont
  {D.}~\bibnamefont {DiFiore}}, \bibinfo {author} {\bibfnamefont {D.}~\bibnamefont {Milathianaki}}, \bibinfo {author} {\bibfnamefont {A.~R.}\ \bibnamefont {Fry}}, \bibinfo {author} {\bibfnamefont {A.}~\bibnamefont {Miahnahri}}, \bibinfo {author} {\bibfnamefont {D.~W.}\ \bibnamefont {Schafer}}, \bibinfo {author} {\bibfnamefont {M.}~\bibnamefont {Messerschmidt}}, \bibinfo {author} {\bibfnamefont {M.~M.}\ \bibnamefont {Seibert}}, \bibinfo {author} {\bibfnamefont {J.~E.}\ \bibnamefont {Koglin}}, \bibinfo {author} {\bibfnamefont {D.}~\bibnamefont {Sokaras}}, \bibinfo {author} {\bibfnamefont {T.-C.}\ \bibnamefont {Weng}}, \bibinfo {author} {\bibfnamefont {J.}~\bibnamefont {Sellberg}}, \bibinfo {author} {\bibfnamefont {M.~J.}\ \bibnamefont {Latimer}}, \bibinfo {author} {\bibfnamefont {R.~W.}\ \bibnamefont {Grosse-Kunstleve}}, \bibinfo {author} {\bibfnamefont {P.~H.}\ \bibnamefont {Zwart}}, \bibinfo {author} {\bibfnamefont {W.~E.}\ \bibnamefont {White}}, \bibinfo {author} {\bibfnamefont {P.}~\bibnamefont {Glatzel}},
  \bibinfo {author} {\bibfnamefont {P.~D.}\ \bibnamefont {Adams}}, \bibinfo {author} {\bibfnamefont {M.~J.}\ \bibnamefont {Bogan}}, \bibinfo {author} {\bibfnamefont {G.~J.}\ \bibnamefont {Williams}}, \bibinfo {author} {\bibfnamefont {S.}~\bibnamefont {Boutet}}, \bibinfo {author} {\bibfnamefont {J.}~\bibnamefont {Messinger}}, \bibinfo {author} {\bibfnamefont {A.}~\bibnamefont {Zouni}}, \bibinfo {author} {\bibfnamefont {N.~K.}\ \bibnamefont {Sauter}}, \bibinfo {author} {\bibfnamefont {V.~K.}\ \bibnamefont {Yachandra}}, \bibinfo {author} {\bibfnamefont {U.}~\bibnamefont {Bergmann}},\ and\ \bibinfo {author} {\bibfnamefont {J.}~\bibnamefont {Yano}},\ }\href@noop {} {\bibfield  {journal} {\bibinfo  {journal} {Science}\ }\textbf {\bibinfo {volume} {340}},\ \bibinfo {pages} {491} (\bibinfo {year} {2013})}\BibitemShut {NoStop}%
\bibitem [{\citenamefont {Reinhard}\ \emph {et~al.}(2021)\citenamefont {Reinhard}, \citenamefont {Mara}, \citenamefont {Kroll}, \citenamefont {Lim}, \citenamefont {Hadt}, \citenamefont {Alonso-Mori}, \citenamefont {Chollet}, \citenamefont {Glownia}, \citenamefont {Nelson}, \citenamefont {Sokaras}, \citenamefont {Kunnus}, \citenamefont {Driel}, \citenamefont {Hartsock}, \citenamefont {Kjaer}, \citenamefont {Weninger}, \citenamefont {Biasin}, \citenamefont {Gee}, \citenamefont {Hodgson}, \citenamefont {Hedman}, \citenamefont {Bergmann}, \citenamefont {Solomon},\ and\ \citenamefont {Gaffney}}]{Reinhard_NatureComm21}%
  \BibitemOpen
  \bibfield  {author} {\bibinfo {author} {\bibfnamefont {M.~E.}\ \bibnamefont {Reinhard}}, \bibinfo {author} {\bibfnamefont {M.~W.}\ \bibnamefont {Mara}}, \bibinfo {author} {\bibfnamefont {T.}~\bibnamefont {Kroll}}, \bibinfo {author} {\bibfnamefont {H.}~\bibnamefont {Lim}}, \bibinfo {author} {\bibfnamefont {R.~G.}\ \bibnamefont {Hadt}}, \bibinfo {author} {\bibfnamefont {R.}~\bibnamefont {Alonso-Mori}}, \bibinfo {author} {\bibfnamefont {M.}~\bibnamefont {Chollet}}, \bibinfo {author} {\bibfnamefont {J.~M.}\ \bibnamefont {Glownia}}, \bibinfo {author} {\bibfnamefont {S.}~\bibnamefont {Nelson}}, \bibinfo {author} {\bibfnamefont {D.}~\bibnamefont {Sokaras}}, \bibinfo {author} {\bibfnamefont {K.}~\bibnamefont {Kunnus}}, \bibinfo {author} {\bibfnamefont {T.~B.~v.}\ \bibnamefont {Driel}}, \bibinfo {author} {\bibfnamefont {R.~W.}\ \bibnamefont {Hartsock}}, \bibinfo {author} {\bibfnamefont {K.~S.}\ \bibnamefont {Kjaer}}, \bibinfo {author} {\bibfnamefont {C.}~\bibnamefont {Weninger}}, \bibinfo {author} {\bibfnamefont
  {E.}~\bibnamefont {Biasin}}, \bibinfo {author} {\bibfnamefont {L.~B.}\ \bibnamefont {Gee}}, \bibinfo {author} {\bibfnamefont {K.~O.}\ \bibnamefont {Hodgson}}, \bibinfo {author} {\bibfnamefont {B.}~\bibnamefont {Hedman}}, \bibinfo {author} {\bibfnamefont {U.}~\bibnamefont {Bergmann}}, \bibinfo {author} {\bibfnamefont {E.~I.}\ \bibnamefont {Solomon}},\ and\ \bibinfo {author} {\bibfnamefont {K.~J.}\ \bibnamefont {Gaffney}},\ }\href@noop {} {\bibfield  {journal} {\bibinfo  {journal} {Nature Communications}\ }\textbf {\bibinfo {volume} {12}},\ \bibinfo {pages} {1086} (\bibinfo {year} {2021})}\BibitemShut {NoStop}%
\bibitem [{\citenamefont {Messinger}\ \emph {et~al.}(2001)\citenamefont {Messinger}, \citenamefont {Robblee}, \citenamefont {Bergmann}, \citenamefont {Fernandez}, \citenamefont {Glatzel}, \citenamefont {Visser}, \citenamefont {Cinco}, \citenamefont {McFarlane}, \citenamefont {Bellacchio}, \citenamefont {Pizarro}, \citenamefont {Cramer}, \citenamefont {Sauer}, \citenamefont {Klein},\ and\ \citenamefont {Yachandra}}]{Messinger_JACS01}%
  \BibitemOpen
  \bibfield  {author} {\bibinfo {author} {\bibfnamefont {J.}~\bibnamefont {Messinger}}, \bibinfo {author} {\bibfnamefont {J.~H.}\ \bibnamefont {Robblee}}, \bibinfo {author} {\bibfnamefont {U.}~\bibnamefont {Bergmann}}, \bibinfo {author} {\bibfnamefont {C.}~\bibnamefont {Fernandez}}, \bibinfo {author} {\bibfnamefont {P.}~\bibnamefont {Glatzel}}, \bibinfo {author} {\bibfnamefont {H.}~\bibnamefont {Visser}}, \bibinfo {author} {\bibfnamefont {R.~M.}\ \bibnamefont {Cinco}}, \bibinfo {author} {\bibfnamefont {K.~L.}\ \bibnamefont {McFarlane}}, \bibinfo {author} {\bibfnamefont {E.}~\bibnamefont {Bellacchio}}, \bibinfo {author} {\bibfnamefont {S.~A.}\ \bibnamefont {Pizarro}}, \bibinfo {author} {\bibfnamefont {S.~P.}\ \bibnamefont {Cramer}}, \bibinfo {author} {\bibfnamefont {K.}~\bibnamefont {Sauer}}, \bibinfo {author} {\bibfnamefont {M.~P.}\ \bibnamefont {Klein}},\ and\ \bibinfo {author} {\bibfnamefont {V.~K.}\ \bibnamefont {Yachandra}},\ }\href@noop {} {\bibfield  {journal} {\bibinfo  {journal} {Journal of the
  American Chemical Society}\ }\textbf {\bibinfo {volume} {123}},\ \bibinfo {pages} {7804} (\bibinfo {year} {2001})}\BibitemShut {NoStop}%
\bibitem [{\citenamefont {Yoneda}\ \emph {et~al.}(2015)\citenamefont {Yoneda}, \citenamefont {Inubushi}, \citenamefont {Nagamine}, \citenamefont {Michine}, \citenamefont {Ohashi}, \citenamefont {Yumoto}, \citenamefont {Yamauchi}, \citenamefont {Mimura}, \citenamefont {Kitamura}, \citenamefont {Katayama}, \citenamefont {Ishikawa},\ and\ \citenamefont {Yabashi}}]{Yoneda_Atomic_2015}%
  \BibitemOpen
  \bibfield  {author} {\bibinfo {author} {\bibfnamefont {H.}~\bibnamefont {Yoneda}}, \bibinfo {author} {\bibfnamefont {Y.}~\bibnamefont {Inubushi}}, \bibinfo {author} {\bibfnamefont {K.}~\bibnamefont {Nagamine}}, \bibinfo {author} {\bibfnamefont {Y.}~\bibnamefont {Michine}}, \bibinfo {author} {\bibfnamefont {H.}~\bibnamefont {Ohashi}}, \bibinfo {author} {\bibfnamefont {H.}~\bibnamefont {Yumoto}}, \bibinfo {author} {\bibfnamefont {K.}~\bibnamefont {Yamauchi}}, \bibinfo {author} {\bibfnamefont {H.}~\bibnamefont {Mimura}}, \bibinfo {author} {\bibfnamefont {H.}~\bibnamefont {Kitamura}}, \bibinfo {author} {\bibfnamefont {T.}~\bibnamefont {Katayama}}, \bibinfo {author} {\bibfnamefont {T.}~\bibnamefont {Ishikawa}},\ and\ \bibinfo {author} {\bibfnamefont {M.}~\bibnamefont {Yabashi}},\ }\href {https://doi.org/10.1038/nature14894} {\bibfield  {journal} {\bibinfo  {journal} {Nature}\ }\textbf {\bibinfo {volume} {524}},\ \bibinfo {pages} {446} (\bibinfo {year} {2015})}\BibitemShut {NoStop}%
\bibitem [{\citenamefont {Kroll}\ \emph {et~al.}(2020)\citenamefont {Kroll}, \citenamefont {Weninger}, \citenamefont {Fuller}, \citenamefont {Guetg}, \citenamefont {Benediktovitch}, \citenamefont {Zhang}, \citenamefont {Marinelli}, \citenamefont {Alonso-Mori}, \citenamefont {Aquila}, \citenamefont {Liang}, \citenamefont {Koglin}, \citenamefont {Koralek}, \citenamefont {Sokaras}, \citenamefont {Zhu}, \citenamefont {Kern}, \citenamefont {Yano}, \citenamefont {Yachandra}, \citenamefont {Rohringer}, \citenamefont {Lutman},\ and\ \citenamefont {Bergmann}}]{Kroll_PRL20}%
  \BibitemOpen
  \bibfield  {author} {\bibinfo {author} {\bibfnamefont {T.}~\bibnamefont {Kroll}}, \bibinfo {author} {\bibfnamefont {C.}~\bibnamefont {Weninger}}, \bibinfo {author} {\bibfnamefont {F.~D.}\ \bibnamefont {Fuller}}, \bibinfo {author} {\bibfnamefont {M.~W.}\ \bibnamefont {Guetg}}, \bibinfo {author} {\bibfnamefont {A.}~\bibnamefont {Benediktovitch}}, \bibinfo {author} {\bibfnamefont {Y.}~\bibnamefont {Zhang}}, \bibinfo {author} {\bibfnamefont {A.}~\bibnamefont {Marinelli}}, \bibinfo {author} {\bibfnamefont {R.}~\bibnamefont {Alonso-Mori}}, \bibinfo {author} {\bibfnamefont {A.}~\bibnamefont {Aquila}}, \bibinfo {author} {\bibfnamefont {M.}~\bibnamefont {Liang}}, \bibinfo {author} {\bibfnamefont {J.~E.}\ \bibnamefont {Koglin}}, \bibinfo {author} {\bibfnamefont {J.}~\bibnamefont {Koralek}}, \bibinfo {author} {\bibfnamefont {D.}~\bibnamefont {Sokaras}}, \bibinfo {author} {\bibfnamefont {D.}~\bibnamefont {Zhu}}, \bibinfo {author} {\bibfnamefont {J.}~\bibnamefont {Kern}}, \bibinfo {author} {\bibfnamefont
  {J.}~\bibnamefont {Yano}}, \bibinfo {author} {\bibfnamefont {V.~K.}\ \bibnamefont {Yachandra}}, \bibinfo {author} {\bibfnamefont {N.}~\bibnamefont {Rohringer}}, \bibinfo {author} {\bibfnamefont {A.}~\bibnamefont {Lutman}},\ and\ \bibinfo {author} {\bibfnamefont {U.}~\bibnamefont {Bergmann}},\ }\href@noop {} {\bibfield  {journal} {\bibinfo  {journal} {Phys. Rev. Lett.}\ }\textbf {\bibinfo {volume} {125}},\ \bibinfo {pages} {037404} (\bibinfo {year} {2020})}\BibitemShut {NoStop}%
\bibitem [{\citenamefont {Doyle}\ \emph {et~al.}(2023)\citenamefont {Doyle}, \citenamefont {Halavanau}, \citenamefont {Zhang}, \citenamefont {Michine}, \citenamefont {Everts}, \citenamefont {Fuller}, \citenamefont {Alonso-Mori}, \citenamefont {Yabashi}, \citenamefont {Inoue}, \citenamefont {Osaka}, \citenamefont {Yamada}, \citenamefont {Inubushi}, \citenamefont {Hara}, \citenamefont {Kern}, \citenamefont {Yano}, \citenamefont {Yachandra}, \citenamefont {Rohringer}, \citenamefont {Yoneda}, \citenamefont {Kroll}, \citenamefont {Pellegrini},\ and\ \citenamefont {Bergmann}}]{Doyle:23}%
  \BibitemOpen
  \bibfield  {author} {\bibinfo {author} {\bibfnamefont {M.~D.}\ \bibnamefont {Doyle}}, \bibinfo {author} {\bibfnamefont {A.}~\bibnamefont {Halavanau}}, \bibinfo {author} {\bibfnamefont {Y.}~\bibnamefont {Zhang}}, \bibinfo {author} {\bibfnamefont {Y.}~\bibnamefont {Michine}}, \bibinfo {author} {\bibfnamefont {J.}~\bibnamefont {Everts}}, \bibinfo {author} {\bibfnamefont {F.}~\bibnamefont {Fuller}}, \bibinfo {author} {\bibfnamefont {R.}~\bibnamefont {Alonso-Mori}}, \bibinfo {author} {\bibfnamefont {M.}~\bibnamefont {Yabashi}}, \bibinfo {author} {\bibfnamefont {I.}~\bibnamefont {Inoue}}, \bibinfo {author} {\bibfnamefont {T.}~\bibnamefont {Osaka}}, \bibinfo {author} {\bibfnamefont {J.}~\bibnamefont {Yamada}}, \bibinfo {author} {\bibfnamefont {Y.}~\bibnamefont {Inubushi}}, \bibinfo {author} {\bibfnamefont {T.}~\bibnamefont {Hara}}, \bibinfo {author} {\bibfnamefont {J.}~\bibnamefont {Kern}}, \bibinfo {author} {\bibfnamefont {J.}~\bibnamefont {Yano}}, \bibinfo {author} {\bibfnamefont {V.~K.}\ \bibnamefont
  {Yachandra}}, \bibinfo {author} {\bibfnamefont {N.}~\bibnamefont {Rohringer}}, \bibinfo {author} {\bibfnamefont {H.}~\bibnamefont {Yoneda}}, \bibinfo {author} {\bibfnamefont {T.}~\bibnamefont {Kroll}}, \bibinfo {author} {\bibfnamefont {C.}~\bibnamefont {Pellegrini}},\ and\ \bibinfo {author} {\bibfnamefont {U.}~\bibnamefont {Bergmann}},\ }\href {https://doi.org/10.1364/OPTICA.485989} {\bibfield  {journal} {\bibinfo  {journal} {Optica}\ }\textbf {\bibinfo {volume} {10}},\ \bibinfo {pages} {513} (\bibinfo {year} {2023})}\BibitemShut {NoStop}%
\bibitem [{\citenamefont {Emma}\ \emph {et~al.}(2010)\citenamefont {Emma}, \citenamefont {Akre}, \citenamefont {Arthur}, \citenamefont {Bionta}, \citenamefont {Bostedt}, \citenamefont {Bozek}, \citenamefont {Brachmann}, \citenamefont {Bucksbaum}, \citenamefont {Coffee}, \citenamefont {Decker}, \citenamefont {Ding}, \citenamefont {Dowell}, \citenamefont {Edstrom}, \citenamefont {Fisher}, \citenamefont {Frisch}, \citenamefont {Gilevich}, \citenamefont {Hastings}, \citenamefont {Hays}, \citenamefont {Hering}, \citenamefont {Huang}, \citenamefont {Iverson}, \citenamefont {Loos}, \citenamefont {Messerschmidt}, \citenamefont {Miahnahri}, \citenamefont {Moeller}, \citenamefont {Nuhn}, \citenamefont {Pile}, \citenamefont {Ratner}, \citenamefont {Rzepiela}, \citenamefont {Schultz}, \citenamefont {Smith}, \citenamefont {Stefan}, \citenamefont {Tompkins}, \citenamefont {Turner}, \citenamefont {Welch}, \citenamefont {White}, \citenamefont {Wu}, \citenamefont {Yocky},\ and\ \citenamefont {Galayda}}]{Emma_First_2010}%
  \BibitemOpen
  \bibfield  {author} {\bibinfo {author} {\bibfnamefont {P.}~\bibnamefont {Emma}}, \bibinfo {author} {\bibfnamefont {R.}~\bibnamefont {Akre}}, \bibinfo {author} {\bibfnamefont {J.}~\bibnamefont {Arthur}}, \bibinfo {author} {\bibfnamefont {R.}~\bibnamefont {Bionta}}, \bibinfo {author} {\bibfnamefont {C.}~\bibnamefont {Bostedt}}, \bibinfo {author} {\bibfnamefont {J.}~\bibnamefont {Bozek}}, \bibinfo {author} {\bibfnamefont {A.}~\bibnamefont {Brachmann}}, \bibinfo {author} {\bibfnamefont {P.}~\bibnamefont {Bucksbaum}}, \bibinfo {author} {\bibfnamefont {R.}~\bibnamefont {Coffee}}, \bibinfo {author} {\bibfnamefont {F.-J.}\ \bibnamefont {Decker}}, \bibinfo {author} {\bibfnamefont {Y.}~\bibnamefont {Ding}}, \bibinfo {author} {\bibfnamefont {D.}~\bibnamefont {Dowell}}, \bibinfo {author} {\bibfnamefont {S.}~\bibnamefont {Edstrom}}, \bibinfo {author} {\bibfnamefont {A.}~\bibnamefont {Fisher}}, \bibinfo {author} {\bibfnamefont {J.}~\bibnamefont {Frisch}}, \bibinfo {author} {\bibfnamefont {S.}~\bibnamefont {Gilevich}},
  \bibinfo {author} {\bibfnamefont {J.}~\bibnamefont {Hastings}}, \bibinfo {author} {\bibfnamefont {G.}~\bibnamefont {Hays}}, \bibinfo {author} {\bibfnamefont {P.}~\bibnamefont {Hering}}, \bibinfo {author} {\bibfnamefont {Z.}~\bibnamefont {Huang}}, \bibinfo {author} {\bibfnamefont {R.}~\bibnamefont {Iverson}}, \bibinfo {author} {\bibfnamefont {H.}~\bibnamefont {Loos}}, \bibinfo {author} {\bibfnamefont {M.}~\bibnamefont {Messerschmidt}}, \bibinfo {author} {\bibfnamefont {A.}~\bibnamefont {Miahnahri}}, \bibinfo {author} {\bibfnamefont {S.}~\bibnamefont {Moeller}}, \bibinfo {author} {\bibfnamefont {H.-D.}\ \bibnamefont {Nuhn}}, \bibinfo {author} {\bibfnamefont {G.}~\bibnamefont {Pile}}, \bibinfo {author} {\bibfnamefont {D.}~\bibnamefont {Ratner}}, \bibinfo {author} {\bibfnamefont {J.}~\bibnamefont {Rzepiela}}, \bibinfo {author} {\bibfnamefont {D.}~\bibnamefont {Schultz}}, \bibinfo {author} {\bibfnamefont {T.}~\bibnamefont {Smith}}, \bibinfo {author} {\bibfnamefont {P.}~\bibnamefont {Stefan}}, \bibinfo {author}
  {\bibfnamefont {H.}~\bibnamefont {Tompkins}}, \bibinfo {author} {\bibfnamefont {J.}~\bibnamefont {Turner}}, \bibinfo {author} {\bibfnamefont {J.}~\bibnamefont {Welch}}, \bibinfo {author} {\bibfnamefont {W.}~\bibnamefont {White}}, \bibinfo {author} {\bibfnamefont {J.}~\bibnamefont {Wu}}, \bibinfo {author} {\bibfnamefont {G.}~\bibnamefont {Yocky}},\ and\ \bibinfo {author} {\bibfnamefont {J.}~\bibnamefont {Galayda}},\ }\href {https://doi.org/10.1038/nphoton.2010.176} {\bibfield  {journal} {\bibinfo  {journal} {Nat. Photon.}\ }\textbf {\bibinfo {volume} {4}},\ \bibinfo {pages} {641} (\bibinfo {year} {2010})}\BibitemShut {NoStop}%
\bibitem [{\citenamefont {Ishikawa}\ \emph {et~al.}(2012)\citenamefont {Ishikawa}, \citenamefont {Aoyagi}, \citenamefont {Asaka}, \citenamefont {Asano}, \citenamefont {Azumi}, \citenamefont {Bizen}, \citenamefont {Ego}, \citenamefont {Fukami}, \citenamefont {Fukui}, \citenamefont {Furukawa}, \citenamefont {Goto}, \citenamefont {Hanaki}, \citenamefont {Hara}, \citenamefont {Hasegawa}, \citenamefont {Hatsui}, \citenamefont {Higashiya}, \citenamefont {Hirono}, \citenamefont {Hosoda}, \citenamefont {Ishii}, \citenamefont {Inagaki}, \citenamefont {Inubushi}, \citenamefont {Itoga}, \citenamefont {Joti}, \citenamefont {Kago}, \citenamefont {Kameshima}, \citenamefont {Kimura}, \citenamefont {Kirihara}, \citenamefont {Kiyomichi}, \citenamefont {Kobayashi}, \citenamefont {Kondo}, \citenamefont {Kudo}, \citenamefont {Maesaka}, \citenamefont {Mar{\'e}chal}, \citenamefont {Masuda}, \citenamefont {Matsubara}, \citenamefont {Matsumoto}, \citenamefont {Matsushita}, \citenamefont {Matsui}, \citenamefont {Nagasono},
  \citenamefont {Nariyama}, \citenamefont {Ohashi}, \citenamefont {Ohata}, \citenamefont {Ohshima}, \citenamefont {Ono}, \citenamefont {Otake}, \citenamefont {Saji}, \citenamefont {Sakurai}, \citenamefont {Sato}, \citenamefont {Sawada}, \citenamefont {Seike}, \citenamefont {Shirasawa}, \citenamefont {Sugimoto}, \citenamefont {Suzuki}, \citenamefont {Takahashi}, \citenamefont {Takebe}, \citenamefont {Takeshita}, \citenamefont {Tamasaku}, \citenamefont {Tanaka}, \citenamefont {Tanaka}, \citenamefont {Tanaka}, \citenamefont {Togashi}, \citenamefont {Togawa}, \citenamefont {Tokuhisa}, \citenamefont {Tomizawa}, \citenamefont {Tono}, \citenamefont {Wu}, \citenamefont {Yabashi}, \citenamefont {Yamaga}, \citenamefont {Yamashita}, \citenamefont {Yanagida}, \citenamefont {Zhang}, \citenamefont {Shintake}, \citenamefont {Kitamura},\ and\ \citenamefont {Kumagai}}]{Ishikawa_compact_2012}%
  \BibitemOpen
  \bibfield  {author} {\bibinfo {author} {\bibfnamefont {T.}~\bibnamefont {Ishikawa}}, \bibinfo {author} {\bibfnamefont {H.}~\bibnamefont {Aoyagi}}, \bibinfo {author} {\bibfnamefont {T.}~\bibnamefont {Asaka}}, \bibinfo {author} {\bibfnamefont {Y.}~\bibnamefont {Asano}}, \bibinfo {author} {\bibfnamefont {N.}~\bibnamefont {Azumi}}, \bibinfo {author} {\bibfnamefont {T.}~\bibnamefont {Bizen}}, \bibinfo {author} {\bibfnamefont {H.}~\bibnamefont {Ego}}, \bibinfo {author} {\bibfnamefont {K.}~\bibnamefont {Fukami}}, \bibinfo {author} {\bibfnamefont {T.}~\bibnamefont {Fukui}}, \bibinfo {author} {\bibfnamefont {Y.}~\bibnamefont {Furukawa}}, \bibinfo {author} {\bibfnamefont {S.}~\bibnamefont {Goto}}, \bibinfo {author} {\bibfnamefont {H.}~\bibnamefont {Hanaki}}, \bibinfo {author} {\bibfnamefont {T.}~\bibnamefont {Hara}}, \bibinfo {author} {\bibfnamefont {T.}~\bibnamefont {Hasegawa}}, \bibinfo {author} {\bibfnamefont {T.}~\bibnamefont {Hatsui}}, \bibinfo {author} {\bibfnamefont {A.}~\bibnamefont {Higashiya}}, \bibinfo
  {author} {\bibfnamefont {T.}~\bibnamefont {Hirono}}, \bibinfo {author} {\bibfnamefont {N.}~\bibnamefont {Hosoda}}, \bibinfo {author} {\bibfnamefont {M.}~\bibnamefont {Ishii}}, \bibinfo {author} {\bibfnamefont {T.}~\bibnamefont {Inagaki}}, \bibinfo {author} {\bibfnamefont {Y.}~\bibnamefont {Inubushi}}, \bibinfo {author} {\bibfnamefont {T.}~\bibnamefont {Itoga}}, \bibinfo {author} {\bibfnamefont {Y.}~\bibnamefont {Joti}}, \bibinfo {author} {\bibfnamefont {M.}~\bibnamefont {Kago}}, \bibinfo {author} {\bibfnamefont {T.}~\bibnamefont {Kameshima}}, \bibinfo {author} {\bibfnamefont {H.}~\bibnamefont {Kimura}}, \bibinfo {author} {\bibfnamefont {Y.}~\bibnamefont {Kirihara}}, \bibinfo {author} {\bibfnamefont {A.}~\bibnamefont {Kiyomichi}}, \bibinfo {author} {\bibfnamefont {T.}~\bibnamefont {Kobayashi}}, \bibinfo {author} {\bibfnamefont {C.}~\bibnamefont {Kondo}}, \bibinfo {author} {\bibfnamefont {T.}~\bibnamefont {Kudo}}, \bibinfo {author} {\bibfnamefont {H.}~\bibnamefont {Maesaka}}, \bibinfo {author} {\bibfnamefont
  {X.~M.}\ \bibnamefont {Mar{\'e}chal}}, \bibinfo {author} {\bibfnamefont {T.}~\bibnamefont {Masuda}}, \bibinfo {author} {\bibfnamefont {S.}~\bibnamefont {Matsubara}}, \bibinfo {author} {\bibfnamefont {T.}~\bibnamefont {Matsumoto}}, \bibinfo {author} {\bibfnamefont {T.}~\bibnamefont {Matsushita}}, \bibinfo {author} {\bibfnamefont {S.}~\bibnamefont {Matsui}}, \bibinfo {author} {\bibfnamefont {M.}~\bibnamefont {Nagasono}}, \bibinfo {author} {\bibfnamefont {N.}~\bibnamefont {Nariyama}}, \bibinfo {author} {\bibfnamefont {H.}~\bibnamefont {Ohashi}}, \bibinfo {author} {\bibfnamefont {T.}~\bibnamefont {Ohata}}, \bibinfo {author} {\bibfnamefont {T.}~\bibnamefont {Ohshima}}, \bibinfo {author} {\bibfnamefont {S.}~\bibnamefont {Ono}}, \bibinfo {author} {\bibfnamefont {Y.}~\bibnamefont {Otake}}, \bibinfo {author} {\bibfnamefont {C.}~\bibnamefont {Saji}}, \bibinfo {author} {\bibfnamefont {T.}~\bibnamefont {Sakurai}}, \bibinfo {author} {\bibfnamefont {T.}~\bibnamefont {Sato}}, \bibinfo {author} {\bibfnamefont
  {K.}~\bibnamefont {Sawada}}, \bibinfo {author} {\bibfnamefont {T.}~\bibnamefont {Seike}}, \bibinfo {author} {\bibfnamefont {K.}~\bibnamefont {Shirasawa}}, \bibinfo {author} {\bibfnamefont {T.}~\bibnamefont {Sugimoto}}, \bibinfo {author} {\bibfnamefont {S.}~\bibnamefont {Suzuki}}, \bibinfo {author} {\bibfnamefont {S.}~\bibnamefont {Takahashi}}, \bibinfo {author} {\bibfnamefont {H.}~\bibnamefont {Takebe}}, \bibinfo {author} {\bibfnamefont {K.}~\bibnamefont {Takeshita}}, \bibinfo {author} {\bibfnamefont {K.}~\bibnamefont {Tamasaku}}, \bibinfo {author} {\bibfnamefont {H.}~\bibnamefont {Tanaka}}, \bibinfo {author} {\bibfnamefont {R.}~\bibnamefont {Tanaka}}, \bibinfo {author} {\bibfnamefont {T.}~\bibnamefont {Tanaka}}, \bibinfo {author} {\bibfnamefont {T.}~\bibnamefont {Togashi}}, \bibinfo {author} {\bibfnamefont {K.}~\bibnamefont {Togawa}}, \bibinfo {author} {\bibfnamefont {A.}~\bibnamefont {Tokuhisa}}, \bibinfo {author} {\bibfnamefont {H.}~\bibnamefont {Tomizawa}}, \bibinfo {author} {\bibfnamefont
  {K.}~\bibnamefont {Tono}}, \bibinfo {author} {\bibfnamefont {S.}~\bibnamefont {Wu}}, \bibinfo {author} {\bibfnamefont {M.}~\bibnamefont {Yabashi}}, \bibinfo {author} {\bibfnamefont {M.}~\bibnamefont {Yamaga}}, \bibinfo {author} {\bibfnamefont {A.}~\bibnamefont {Yamashita}}, \bibinfo {author} {\bibfnamefont {K.}~\bibnamefont {Yanagida}}, \bibinfo {author} {\bibfnamefont {C.}~\bibnamefont {Zhang}}, \bibinfo {author} {\bibfnamefont {T.}~\bibnamefont {Shintake}}, \bibinfo {author} {\bibfnamefont {H.}~\bibnamefont {Kitamura}},\ and\ \bibinfo {author} {\bibfnamefont {N.}~\bibnamefont {Kumagai}},\ }\href {https://doi.org/10.1038/nphoton.2012.141} {\bibfield  {journal} {\bibinfo  {journal} {Nat. Photon.}\ }\textbf {\bibinfo {volume} {6}},\ \bibinfo {pages} {540} (\bibinfo {year} {2012})}\BibitemShut {NoStop}%
\bibitem [{\citenamefont {Rohringer}\ \emph {et~al.}(2012)\citenamefont {Rohringer}, \citenamefont {Ryan}, \citenamefont {London}, \citenamefont {Purvis}, \citenamefont {Albert}, \citenamefont {Dunn}, \citenamefont {Bozek}, \citenamefont {Bostedt}, \citenamefont {Graf}, \citenamefont {Hill}, \citenamefont {Hau-Riege},\ and\ \citenamefont {Rocca}}]{Rohringer_Atomic_2012}%
  \BibitemOpen
  \bibfield  {author} {\bibinfo {author} {\bibfnamefont {N.}~\bibnamefont {Rohringer}}, \bibinfo {author} {\bibfnamefont {D.}~\bibnamefont {Ryan}}, \bibinfo {author} {\bibfnamefont {R.~A.}\ \bibnamefont {London}}, \bibinfo {author} {\bibfnamefont {M.}~\bibnamefont {Purvis}}, \bibinfo {author} {\bibfnamefont {F.}~\bibnamefont {Albert}}, \bibinfo {author} {\bibfnamefont {J.}~\bibnamefont {Dunn}}, \bibinfo {author} {\bibfnamefont {J.~D.}\ \bibnamefont {Bozek}}, \bibinfo {author} {\bibfnamefont {C.}~\bibnamefont {Bostedt}}, \bibinfo {author} {\bibfnamefont {A.}~\bibnamefont {Graf}}, \bibinfo {author} {\bibfnamefont {R.}~\bibnamefont {Hill}}, \bibinfo {author} {\bibfnamefont {S.~P.}\ \bibnamefont {Hau-Riege}},\ and\ \bibinfo {author} {\bibfnamefont {J.~J.}\ \bibnamefont {Rocca}},\ }\href {https://doi.org/10.1038/nature10721} {\bibfield  {journal} {\bibinfo  {journal} {Nature}\ }\textbf {\bibinfo {volume} {481}},\ \bibinfo {pages} {488} (\bibinfo {year} {2012})}\BibitemShut {NoStop}%
\bibitem [{\citenamefont {Beye}\ \emph {et~al.}(2013)\citenamefont {Beye}, \citenamefont {Schreck}, \citenamefont {Sorgenfrei}, \citenamefont {Trabant}, \citenamefont {Pontius}, \citenamefont {Sch{\"u}\ss{}ler-Langeheine}, \citenamefont {Wurth},\ and\ \citenamefont {F{\"o}hlisch}}]{Beye_Stimulated_2013}%
  \BibitemOpen
  \bibfield  {author} {\bibinfo {author} {\bibfnamefont {M.}~\bibnamefont {Beye}}, \bibinfo {author} {\bibfnamefont {S.}~\bibnamefont {Schreck}}, \bibinfo {author} {\bibfnamefont {F.}~\bibnamefont {Sorgenfrei}}, \bibinfo {author} {\bibfnamefont {C.}~\bibnamefont {Trabant}}, \bibinfo {author} {\bibfnamefont {N.}~\bibnamefont {Pontius}}, \bibinfo {author} {\bibfnamefont {C.}~\bibnamefont {Sch{\"u}\ss{}ler-Langeheine}}, \bibinfo {author} {\bibfnamefont {W.}~\bibnamefont {Wurth}},\ and\ \bibinfo {author} {\bibfnamefont {A.}~\bibnamefont {F{\"o}hlisch}},\ }\href {https://doi.org/10.1038/nature12449} {\bibfield  {journal} {\bibinfo  {journal} {Nature}\ }\textbf {\bibinfo {volume} {501}},\ \bibinfo {pages} {191} (\bibinfo {year} {2013})}\BibitemShut {NoStop}%
\bibitem [{\citenamefont {Kroll}\ \emph {et~al.}(2018)\citenamefont {Kroll}, \citenamefont {Weninger}, \citenamefont {Alonso-Mori}, \citenamefont {Sokaras}, \citenamefont {Zhu}, \citenamefont {Mercadier}, \citenamefont {Majety}, \citenamefont {Marinelli}, \citenamefont {Lutman}, \citenamefont {Guetg}, \citenamefont {Decker}, \citenamefont {Boutet}, \citenamefont {Aquila}, \citenamefont {Koglin}, \citenamefont {Koralek}, \citenamefont {DePonte}, \citenamefont {Kern}, \citenamefont {Fuller}, \citenamefont {Pastor}, \citenamefont {Fransson}, \citenamefont {Zhang}, \citenamefont {Yano}, \citenamefont {Yachandra}, \citenamefont {Rohringer},\ and\ \citenamefont {Bergmann}}]{Kroll_PRL18}%
  \BibitemOpen
  \bibfield  {author} {\bibinfo {author} {\bibfnamefont {T.}~\bibnamefont {Kroll}}, \bibinfo {author} {\bibfnamefont {C.}~\bibnamefont {Weninger}}, \bibinfo {author} {\bibfnamefont {R.}~\bibnamefont {Alonso-Mori}}, \bibinfo {author} {\bibfnamefont {D.}~\bibnamefont {Sokaras}}, \bibinfo {author} {\bibfnamefont {D.}~\bibnamefont {Zhu}}, \bibinfo {author} {\bibfnamefont {L.}~\bibnamefont {Mercadier}}, \bibinfo {author} {\bibfnamefont {V.~P.}\ \bibnamefont {Majety}}, \bibinfo {author} {\bibfnamefont {A.}~\bibnamefont {Marinelli}}, \bibinfo {author} {\bibfnamefont {A.}~\bibnamefont {Lutman}}, \bibinfo {author} {\bibfnamefont {M.~W.}\ \bibnamefont {Guetg}}, \bibinfo {author} {\bibfnamefont {F.-J.}\ \bibnamefont {Decker}}, \bibinfo {author} {\bibfnamefont {S.}~\bibnamefont {Boutet}}, \bibinfo {author} {\bibfnamefont {A.}~\bibnamefont {Aquila}}, \bibinfo {author} {\bibfnamefont {J.}~\bibnamefont {Koglin}}, \bibinfo {author} {\bibfnamefont {J.}~\bibnamefont {Koralek}}, \bibinfo {author} {\bibfnamefont {D.~P.}\
  \bibnamefont {DePonte}}, \bibinfo {author} {\bibfnamefont {J.}~\bibnamefont {Kern}}, \bibinfo {author} {\bibfnamefont {F.~D.}\ \bibnamefont {Fuller}}, \bibinfo {author} {\bibfnamefont {E.}~\bibnamefont {Pastor}}, \bibinfo {author} {\bibfnamefont {T.}~\bibnamefont {Fransson}}, \bibinfo {author} {\bibfnamefont {Y.}~\bibnamefont {Zhang}}, \bibinfo {author} {\bibfnamefont {J.}~\bibnamefont {Yano}}, \bibinfo {author} {\bibfnamefont {V.~K.}\ \bibnamefont {Yachandra}}, \bibinfo {author} {\bibfnamefont {N.}~\bibnamefont {Rohringer}},\ and\ \bibinfo {author} {\bibfnamefont {U.}~\bibnamefont {Bergmann}},\ }\href@noop {} {\bibfield  {journal} {\bibinfo  {journal} {Phys. Rev. Lett.}\ }\textbf {\bibinfo {volume} {120}},\ \bibinfo {pages} {133203} (\bibinfo {year} {2018})}\BibitemShut {NoStop}%
\bibitem [{\citenamefont {Benediktovitch}\ \emph {et~al.}(2019)\citenamefont {Benediktovitch}, \citenamefont {Majety},\ and\ \citenamefont {Rohringer}}]{Benediktovitch_PRA19}%
  \BibitemOpen
  \bibfield  {author} {\bibinfo {author} {\bibfnamefont {A.}~\bibnamefont {Benediktovitch}}, \bibinfo {author} {\bibfnamefont {V.~P.}\ \bibnamefont {Majety}},\ and\ \bibinfo {author} {\bibfnamefont {N.}~\bibnamefont {Rohringer}},\ }\href@noop {} {\bibfield  {journal} {\bibinfo  {journal} {Phys. Rev. A}\ }\textbf {\bibinfo {volume} {99}},\ \bibinfo {pages} {013839} (\bibinfo {year} {2019})}\BibitemShut {NoStop}%
\bibitem [{\citenamefont {Mercadier}\ \emph {et~al.}(2019)\citenamefont {Mercadier}, \citenamefont {Benediktovitch}, \citenamefont {Weninger}, \citenamefont {Blessenohl}, \citenamefont {Bernitt}, \citenamefont {Bekker}, \citenamefont {Dobrodey}, \citenamefont {Sanchez-Gonzalez}, \citenamefont {Erk}, \citenamefont {Bomme}, \citenamefont {Boll}, \citenamefont {Yin}, \citenamefont {Majety}, \citenamefont {Steinbr{\"u}gge}, \citenamefont {Khalal}, \citenamefont {Penent}, \citenamefont {Palaudoux}, \citenamefont {Lablanquie}, \citenamefont {Rudenko}, \citenamefont {Rolles}, \citenamefont {Lopez-Urrutia},\ and\ \citenamefont {Rohringer}}]{Mercadier_PRL19}%
  \BibitemOpen
  \bibfield  {author} {\bibinfo {author} {\bibfnamefont {L.}~\bibnamefont {Mercadier}}, \bibinfo {author} {\bibfnamefont {A.}~\bibnamefont {Benediktovitch}}, \bibinfo {author} {\bibfnamefont {C.}~\bibnamefont {Weninger}}, \bibinfo {author} {\bibfnamefont {M.~A.}\ \bibnamefont {Blessenohl}}, \bibinfo {author} {\bibfnamefont {S.}~\bibnamefont {Bernitt}}, \bibinfo {author} {\bibfnamefont {H.}~\bibnamefont {Bekker}}, \bibinfo {author} {\bibfnamefont {S.}~\bibnamefont {Dobrodey}}, \bibinfo {author} {\bibfnamefont {A.}~\bibnamefont {Sanchez-Gonzalez}}, \bibinfo {author} {\bibfnamefont {B.}~\bibnamefont {Erk}}, \bibinfo {author} {\bibfnamefont {C.}~\bibnamefont {Bomme}}, \bibinfo {author} {\bibfnamefont {R.}~\bibnamefont {Boll}}, \bibinfo {author} {\bibfnamefont {Z.}~\bibnamefont {Yin}}, \bibinfo {author} {\bibfnamefont {V.~P.}\ \bibnamefont {Majety}}, \bibinfo {author} {\bibfnamefont {R.}~\bibnamefont {Steinbr{\"u}gge}}, \bibinfo {author} {\bibfnamefont {M.~A.}\ \bibnamefont {Khalal}}, \bibinfo {author}
  {\bibfnamefont {F.}~\bibnamefont {Penent}}, \bibinfo {author} {\bibfnamefont {J.}~\bibnamefont {Palaudoux}}, \bibinfo {author} {\bibfnamefont {P.}~\bibnamefont {Lablanquie}}, \bibinfo {author} {\bibfnamefont {A.}~\bibnamefont {Rudenko}}, \bibinfo {author} {\bibfnamefont {D.}~\bibnamefont {Rolles}}, \bibinfo {author} {\bibfnamefont {J.~R.~C.}\ \bibnamefont {Lopez-Urrutia}},\ and\ \bibinfo {author} {\bibfnamefont {N.}~\bibnamefont {Rohringer}},\ }\href@noop {} {\bibfield  {journal} {\bibinfo  {journal} {Phys. Rev. Lett.}\ }\textbf {\bibinfo {volume} {120}},\ \bibinfo {pages} {023201} (\bibinfo {year} {2019})}\BibitemShut {NoStop}%
\bibitem [{\citenamefont {Bergmann}(2024)}]{Bergmann_PR24}%
  \BibitemOpen
  \bibfield  {author} {\bibinfo {author} {\bibfnamefont {U.}~\bibnamefont {Bergmann}},\ }\href@noop {} {\bibfield  {journal} {\bibinfo  {journal} {Photosynthesis Research}\ }\textbf {\bibinfo {volume} {162}},\ \bibinfo {pages} {371} (\bibinfo {year} {2024})}\BibitemShut {NoStop}%
\bibitem [{\citenamefont {Hara}\ \emph {et~al.}(2013)\citenamefont {Hara}, \citenamefont {Inubushi}, \citenamefont {Katayama}, \citenamefont {Sato}, \citenamefont {Tanaka}, \citenamefont {Tanaka}, \citenamefont {Togashi}, \citenamefont {Togawa}, \citenamefont {Tono}, \citenamefont {Yabashi} \emph {et~al.}}]{hara2013two}%
  \BibitemOpen
  \bibfield  {author} {\bibinfo {author} {\bibfnamefont {T.}~\bibnamefont {Hara}}, \bibinfo {author} {\bibfnamefont {Y.}~\bibnamefont {Inubushi}}, \bibinfo {author} {\bibfnamefont {T.}~\bibnamefont {Katayama}}, \bibinfo {author} {\bibfnamefont {T.}~\bibnamefont {Sato}}, \bibinfo {author} {\bibfnamefont {H.}~\bibnamefont {Tanaka}}, \bibinfo {author} {\bibfnamefont {T.}~\bibnamefont {Tanaka}}, \bibinfo {author} {\bibfnamefont {T.}~\bibnamefont {Togashi}}, \bibinfo {author} {\bibfnamefont {K.}~\bibnamefont {Togawa}}, \bibinfo {author} {\bibfnamefont {K.}~\bibnamefont {Tono}}, \bibinfo {author} {\bibfnamefont {M.}~\bibnamefont {Yabashi}}, \emph {et~al.},\ }\href@noop {} {\bibfield  {journal} {\bibinfo  {journal} {Nature communications}\ }\textbf {\bibinfo {volume} {4}},\ \bibinfo {pages} {1} (\bibinfo {year} {2013})}\BibitemShut {NoStop}%
\bibitem [{\citenamefont {Inoue}\ \emph {et~al.}(2020)\citenamefont {Inoue}, \citenamefont {Osaka}, \citenamefont {Hara},\ and\ \citenamefont {Yabashi}}]{Inoue_JSR20}%
  \BibitemOpen
  \bibfield  {author} {\bibinfo {author} {\bibfnamefont {I.}~\bibnamefont {Inoue}}, \bibinfo {author} {\bibfnamefont {T.}~\bibnamefont {Osaka}}, \bibinfo {author} {\bibfnamefont {T.}~\bibnamefont {Hara}},\ and\ \bibinfo {author} {\bibfnamefont {M.}~\bibnamefont {Yabashi}},\ }\href {https://doi.org/10.1107/S1600577520011716} {\bibfield  {journal} {\bibinfo  {journal} {Journal of Synchrotron Radiation}\ }\textbf {\bibinfo {volume} {27}},\ \bibinfo {pages} {1720} (\bibinfo {year} {2020})}\BibitemShut {NoStop}%
\bibitem [{\citenamefont {Krause}\ and\ \citenamefont {Oliver}(1979)}]{Krause_1979}%
  \BibitemOpen
  \bibfield  {author} {\bibinfo {author} {\bibfnamefont {M.~O.}\ \bibnamefont {Krause}}\ and\ \bibinfo {author} {\bibfnamefont {J.~H.}\ \bibnamefont {Oliver}},\ }\href {https://doi.org/10.1063/1.555595} {\bibfield  {journal} {\bibinfo  {journal} {Journal of Physical and Chemical Reference Data}\ }\textbf {\bibinfo {volume} {8}},\ \bibinfo {pages} {329} (\bibinfo {year} {1979})}\BibitemShut {NoStop}%
\bibitem [{\citenamefont {Weakly}\ \emph {et~al.}(2023)\citenamefont {Weakly}, \citenamefont {Liekhus-Schmaltz}, \citenamefont {Poulter}, \citenamefont {Biasin}, \citenamefont {Alonso-Mori}, \citenamefont {Aquila}, \citenamefont {Boutet}, \citenamefont {Fuller}, \citenamefont {Ho}, \citenamefont {Kroll}, \citenamefont {Loe}, \citenamefont {Lutman}, \citenamefont {Zhu}, \citenamefont {Bergmann}, \citenamefont {Schoenlein}, \citenamefont {Govind},\ and\ \citenamefont {Khalil}}]{Weakly_NatureComm23}%
  \BibitemOpen
  \bibfield  {author} {\bibinfo {author} {\bibfnamefont {R.~B.}\ \bibnamefont {Weakly}}, \bibinfo {author} {\bibfnamefont {C.~E.}\ \bibnamefont {Liekhus-Schmaltz}}, \bibinfo {author} {\bibfnamefont {B.~I.}\ \bibnamefont {Poulter}}, \bibinfo {author} {\bibfnamefont {E.}~\bibnamefont {Biasin}}, \bibinfo {author} {\bibfnamefont {R.}~\bibnamefont {Alonso-Mori}}, \bibinfo {author} {\bibfnamefont {A.}~\bibnamefont {Aquila}}, \bibinfo {author} {\bibfnamefont {S.}~\bibnamefont {Boutet}}, \bibinfo {author} {\bibfnamefont {F.~D.}\ \bibnamefont {Fuller}}, \bibinfo {author} {\bibfnamefont {P.~J.}\ \bibnamefont {Ho}}, \bibinfo {author} {\bibfnamefont {T.}~\bibnamefont {Kroll}}, \bibinfo {author} {\bibfnamefont {C.~M.}\ \bibnamefont {Loe}}, \bibinfo {author} {\bibfnamefont {A.}~\bibnamefont {Lutman}}, \bibinfo {author} {\bibfnamefont {D.}~\bibnamefont {Zhu}}, \bibinfo {author} {\bibfnamefont {U.}~\bibnamefont {Bergmann}}, \bibinfo {author} {\bibfnamefont {R.~W.}\ \bibnamefont {Schoenlein}}, \bibinfo {author} {\bibfnamefont
  {N.}~\bibnamefont {Govind}},\ and\ \bibinfo {author} {\bibfnamefont {M.}~\bibnamefont {Khalil}},\ }\href@noop {} {\bibfield  {journal} {\bibinfo  {journal} {Nature Communications}\ }\textbf {\bibinfo {volume} {14}},\ \bibinfo {pages} {3384} (\bibinfo {year} {2023})}\BibitemShut {NoStop}%
\bibitem [{\citenamefont {Fader}(1985)}]{Fader_IEEE85}%
  \BibitemOpen
  \bibfield  {author} {\bibinfo {author} {\bibfnamefont {W.}~\bibnamefont {Fader}},\ }\href {https://doi.org/10.1109/JQE.1985.1072577} {\bibfield  {journal} {\bibinfo  {journal} {IEEE Journal of Quantum Electronics}\ }\textbf {\bibinfo {volume} {21}},\ \bibinfo {pages} {1838} (\bibinfo {year} {1985})}\BibitemShut {NoStop}%
\bibitem [{\citenamefont {A.}(1989)}]{Yaryv_book89}%
  \BibitemOpen
  \bibfield  {author} {\bibinfo {author} {\bibfnamefont {Y.}~\bibnamefont {A.}},\ }\href@noop {} {\emph {\bibinfo {title} {Quantum Electronics}}}\ (\bibinfo  {publisher} {John Wiley \& Sons Inc., NY},\ \bibinfo {year} {1989})\BibitemShut {NoStop}%
\bibitem [{\citenamefont {Allaria}\ \emph {et~al.}(2010)\citenamefont {Allaria}, \citenamefont {Danailov},\ and\ \citenamefont {De~Ninno}}]{Allaria_EPL10}%
  \BibitemOpen
  \bibfield  {author} {\bibinfo {author} {\bibfnamefont {E.}~\bibnamefont {Allaria}}, \bibinfo {author} {\bibfnamefont {M.}~\bibnamefont {Danailov}},\ and\ \bibinfo {author} {\bibfnamefont {G.}~\bibnamefont {De~Ninno}},\ }\href@noop {} {\bibfield  {journal} {\bibinfo  {journal} {Europhysics Letters}\ }\textbf {\bibinfo {volume} {89}},\ \bibinfo {pages} {64005} (\bibinfo {year} {2010})}\BibitemShut {NoStop}%
\bibitem [{\citenamefont {Chuchurka}\ \emph {et~al.}(2024)\citenamefont {Chuchurka}, \citenamefont {Benediktovitch}, \citenamefont {Kru{\v s}i{\v c}}, \citenamefont {Halavanau},\ and\ \citenamefont {Rohringer}}]{Chuchurka_PRA24}%
  \BibitemOpen
  \bibfield  {author} {\bibinfo {author} {\bibfnamefont {S.}~\bibnamefont {Chuchurka}}, \bibinfo {author} {\bibfnamefont {A.}~\bibnamefont {Benediktovitch}}, \bibinfo {author} {\bibfnamefont {{\v S}.}~\bibnamefont {Kru{\v s}i{\v c}}}, \bibinfo {author} {\bibfnamefont {A.}~\bibnamefont {Halavanau}},\ and\ \bibinfo {author} {\bibfnamefont {N.}~\bibnamefont {Rohringer}},\ }\href@noop {} {\bibfield  {journal} {\bibinfo  {journal} {Physical Review A}\ }\textbf {\bibinfo {volume} {109}},\ \bibinfo {pages} {033725} (\bibinfo {year} {2024})}\BibitemShut {NoStop}%
\bibitem [{\citenamefont {Linker}\ \emph {et~al.}(2024)\citenamefont {Linker}, \citenamefont {Halavanau}, \citenamefont {Kroll}, \citenamefont {Benediktovitch}, \citenamefont {Zhang}, \citenamefont {Michine}, \citenamefont {Chuchurka}, \citenamefont {Abhari}, \citenamefont {Ronchetti}, \citenamefont {Fransson}, \citenamefont {Weninger}, \citenamefont {Fuller}, \citenamefont {Aquila}, \citenamefont {Alonso-Mori}, \citenamefont {Boutet}, \citenamefont {Guetg}, \citenamefont {Marinelli}, \citenamefont {Lutman}, \citenamefont {Yabashi}, \citenamefont {Inoue}, \citenamefont {Osaka}, \citenamefont {Yamada}, \citenamefont {Inubushi}, \citenamefont {Yamaguchi}, \citenamefont {Hara}, \citenamefont {Babu}, \citenamefont {Salpekar}, \citenamefont {Sayed}, \citenamefont {Ajayan}, \citenamefont {Kern}, \citenamefont {Yano}, \citenamefont {Yachandra}, \citenamefont {Kling}, \citenamefont {Pellegrini}, \citenamefont {Yoneda}, \citenamefont {Rohringer},\ and\ \citenamefont {Bergmann}}]{Linker_arXiv24}%
  \BibitemOpen
  \bibfield  {author} {\bibinfo {author} {\bibfnamefont {T.~M.}\ \bibnamefont {Linker}}, \bibinfo {author} {\bibfnamefont {A.}~\bibnamefont {Halavanau}}, \bibinfo {author} {\bibfnamefont {T.}~\bibnamefont {Kroll}}, \bibinfo {author} {\bibfnamefont {A.}~\bibnamefont {Benediktovitch}}, \bibinfo {author} {\bibfnamefont {Y.}~\bibnamefont {Zhang}}, \bibinfo {author} {\bibfnamefont {Y.}~\bibnamefont {Michine}}, \bibinfo {author} {\bibfnamefont {S.}~\bibnamefont {Chuchurka}}, \bibinfo {author} {\bibfnamefont {Z.}~\bibnamefont {Abhari}}, \bibinfo {author} {\bibfnamefont {D.}~\bibnamefont {Ronchetti}}, \bibinfo {author} {\bibfnamefont {T.}~\bibnamefont {Fransson}}, \bibinfo {author} {\bibfnamefont {C.}~\bibnamefont {Weninger}}, \bibinfo {author} {\bibfnamefont {F.~D.}\ \bibnamefont {Fuller}}, \bibinfo {author} {\bibfnamefont {A.}~\bibnamefont {Aquila}}, \bibinfo {author} {\bibfnamefont {R.}~\bibnamefont {Alonso-Mori}}, \bibinfo {author} {\bibfnamefont {S.}~\bibnamefont {Boutet}}, \bibinfo {author} {\bibfnamefont
  {M.~W.}\ \bibnamefont {Guetg}}, \bibinfo {author} {\bibfnamefont {A.}~\bibnamefont {Marinelli}}, \bibinfo {author} {\bibfnamefont {A.~A.}\ \bibnamefont {Lutman}}, \bibinfo {author} {\bibfnamefont {M.}~\bibnamefont {Yabashi}}, \bibinfo {author} {\bibfnamefont {I.}~\bibnamefont {Inoue}}, \bibinfo {author} {\bibfnamefont {T.}~\bibnamefont {Osaka}}, \bibinfo {author} {\bibfnamefont {J.}~\bibnamefont {Yamada}}, \bibinfo {author} {\bibfnamefont {Y.}~\bibnamefont {Inubushi}}, \bibinfo {author} {\bibfnamefont {G.}~\bibnamefont {Yamaguchi}}, \bibinfo {author} {\bibfnamefont {T.}~\bibnamefont {Hara}}, \bibinfo {author} {\bibfnamefont {G.}~\bibnamefont {Babu}}, \bibinfo {author} {\bibfnamefont {D.}~\bibnamefont {Salpekar}}, \bibinfo {author} {\bibfnamefont {F.~N.}\ \bibnamefont {Sayed}}, \bibinfo {author} {\bibfnamefont {P.~M.}\ \bibnamefont {Ajayan}}, \bibinfo {author} {\bibfnamefont {J.}~\bibnamefont {Kern}}, \bibinfo {author} {\bibfnamefont {J.}~\bibnamefont {Yano}}, \bibinfo {author} {\bibfnamefont {V.~K.}\
  \bibnamefont {Yachandra}}, \bibinfo {author} {\bibfnamefont {M.~F.}\ \bibnamefont {Kling}}, \bibinfo {author} {\bibfnamefont {C.}~\bibnamefont {Pellegrini}}, \bibinfo {author} {\bibfnamefont {H.}~\bibnamefont {Yoneda}}, \bibinfo {author} {\bibfnamefont {N.}~\bibnamefont {Rohringer}},\ and\ \bibinfo {author} {\bibfnamefont {U.}~\bibnamefont {Bergmann}},\ }\href@noop {} {\bibfield  {journal} {\bibinfo  {journal} {arXiv:2409.06914}\ } (\bibinfo {year} {2024})}\BibitemShut {NoStop}%
\end{thebibliography}%

\end{document}